\documentclass[prd, nofootinbib, preprint, 12 pt]{revtex4}
\usepackage{amsfonts}
\usepackage{amsmath}
\usepackage{amssymb}
\usepackage{epsfig}
\usepackage{subfigure}
\usepackage{graphicx,color}
\usepackage{booktabs}
\usepackage{multirow}
\usepackage{enumerate}
\usepackage{multirow}
\usepackage[dvipdfm,  
            pdfstartview=FitH,
            CJKbookmarks=true,
            bookmarksnumbered=true,
            bookmarksopen=true,
            linkcolor=blue,
            anchorcolor=blue,
            citecolor=blue
            ]{hyperref}
\usepackage{indentfirst}
\allowdisplaybreaks[4]

\begin{document}
\title{$B\to \pi\ell^{+}\ell^{-}$ decays revisited in the standard model}
\author{
Zuo-Hong Li$~^{a}$\footnote{lizh@ytu.edu.cn}, Zong-Guo
Si$~^{b}$\footnote{zgsi@sdu.edu.cn}, Ying
Wang$^{b}$\footnote{wang\_y@mail.sdu.edu.cn} and Nan
Zhu$~^{a}$\footnote{zhunan426@126.com}} \affiliation{{\small\it
$^a$ Department of Physics, Yantai University, Yantai, 264005, P.R.China\\
$^b$ Department of Physics, Shandong University, Jinan, 25100, P. R. China}}
\date{\today}
\begin{abstract}
A new estimate is presented of the dileptonic $B$ decays
$B\to\pi\ell^+\ell^-(\ell=e,\mu,\tau)$ in naive factorization within
the standard-model (SM) framework. Using a combination of several
approaches, we investigate the behavior of the $B\to \pi$ form
factors in the entire region of the momentum transfer squared $q^2$.
For the vector and scalar form factors, we employ the light cone sum
rule (LCSR) with a chiral current correlator to estimate, at twist-2
next-to-leading order (NLO) accuracy, their shapes in small and
intermediate kinematical region. Then a simultaneous fit to a
Bourrely-Caprini-Lellouch (BCL) parametrization is performed of the
sum rule predictions and the corresponding lattice QCD (LQCD)
results available at some high $q^2$'s. The same approach is applied
for the tensor form factor, except that at large $q^2$ we use as
input the LQCD data on the corresponding $B\to K$ form factor in
combination with a $SU_F(3)$ symmetry breaking ansatz. Employing the
fitted BCL parameterizations, we evaluate, as an illustrative
example, several of the observables of the charged decay modes $B^-
\to\pi^-\ell^+\ell^-$, including the dilepton invariant mass
distribution and branching ratio. For the dielectron and dimuon
modes, the branching ratios are estimated at $\mathcal{B}(B^-
\to\pi^-e^+e^-)=(2.263^{+0.227}_{-0.192})\times10^{-8}$ and
$\mathcal{B}(B^-\to
\pi^-\mu^+\mu^-)=(2.259^{+0.226}_{-0.191})\times10^{-8}$. The latter
shows an excellent agreement with the recent experimental
measurement at LHCb and hence puts a stringent constraint on the
contribution from possible new physics. We arrive at, for the ditau
mode, $\mathcal{B}(B^- \to
\pi^-\tau^+\tau^-)=(1.017^{+0.118}_{-0.139})\times10^{-8}$, which is
one order of magnitude larger than the existing theoretical
predictions.
\end{abstract}
\maketitle
\section{Introduction}
Recently, a discovery of the rare decay $B^{+}\to
\pi^{+}\mu^{+}\mu^{-}$ has been reported using a $pp$ collision data
sample, corresponding to integrated luminosity of $1.0 fb^{-1}$,
collected with the LHCb experiment at the Large Hadron Collider
\cite{BPi-Observation}. The branching ratio was measured at
$\mathcal{B}(B^{+}\to
\pi^{+}\mu^{+}\mu^{-})=(2.3\pm0.6(stat.)\pm0.1(syst.))\times10^{-8}$
with $5.2\sigma$ significance. It is the first time flavor-changing
neutral current (FCNC) $b\to d\ell^{+}\ell^{-}$ transitions have
been observed. As data accumulates, more and more attention would be
paid to this aspect.

As is known to all, dileptonic decays of $B$-meson induced by FCNC
$b\to s (d)$ transitions serve as an important avenue to test the
standard model (SM) and search for physics beyond it. The transition
matrix elements contain terms proportional respectively to the
products of the Cabibbo-Kobayashi-Maskawa (CKM) matrix elements,
$V_{tb}V^*_{ts(d)}$, $V_{cb}V^*_{cs(d)}$ and $V_{ub}V^*_{us(d)}$.
Using unitarity of the CKM matrix and neglecting the smaller
$V_{ub}V^*_{us}$ term in comparison to $V_{tb}V^*_{ts}$ and
$V_{cb}V^*_{cs}$, only one independent CKM factor $V_{tb}V^*_{ts}$
is involved in the $b\to s\ell^{+}\ell^{-}$ transitions in this
approximation. In contrast, the CKM factors $V_{tb}V_{td}^{*}$,
$V_{cb}V_{cd}^{*}$ and $V_{ub}V_{ud}^{*}$ are at the same order of
magnitude. Consequently, for the $b\to d$ modes there could be a
considerable CP asymmetry, but meanwhile they are suppressed by a
factor of about $|V_{td}/V_{ts}|$ with respect to the $b\to s $
transitions. This indicates that the $b\to d$ dileptonic transitions
can be complementary to the $b\to s$ ones in probing new physics.

Some theoretical effort has been devoted to research for the
exclusive decays $B\to\pi\ell^{+}\ell^{-}$ ($\ell=e,\mu,\tau$)
within \cite{BPiee-F.Kruger-9706247, BPill-J.J.Wang-0711.0321,
BPill-Z.J.Xiao-1207.0265, BPill-A.Ali-1312.2523,
Bpill-R.N.Faustov-1403.4466, BPill-W.S.Hou-1403.7410} and beyond
\cite{BpillNP-T.M.Aliev-9812272,BpillNP-S.R.Choudhury-0206128,
BpillNP-H.Z.Song-08} the SM. The naive factorization approach is
extensively adopted in these studies and the resulting SM branching
ratios for the dimuonic modes
\cite{BPill-J.J.Wang-0711.0321,BPill-Z.J.Xiao-1207.0265,
BPill-A.Ali-1312.2523,Bpill-R.N.Faustov-1403.4466} turn out to be
consistent with the experimental data. Very recently, in the limits
of the heavy quark mass and large recoil energy, corresponding to
small dilepton mass squared $q^2$, a detailed analysis
\cite{BPill-W.S.Hou-1403.7410} appeared within the framework of QCD
factorization (QCDF)
\cite{M.Beneke-9905312,M.Beneke-0006124,
M.Beneke-0008255,M.Beneke-0106067,A.Ali-0105302,
S.W.Bosch-0106081,M.Beneke-0412400}, which modifies naive
factorization by including the factorizable hard-gluon corrections
to weak vertex and non-factorizable hard spectator scattering
contributions, an important potential source of CP-asymmetry,
$W$-weak annihilation, being identified. A partial understanding is
also achievable of the nonlocal corrections due to soft-gluon
emission and hadronic resonance which could not be covered by QCDF,
using the QCD light cone sum rule (LCSR) approach applied for the
$B\to K$ dileptonic modes \cite{LCSR-K-A.Khodjamirian-1211.0234}.
However, the largest uncertainty in calculating the decay rates and
widths originates from the $B\to\pi $ transition form factors
$f_{+}^{B\to\pi}(q^{2})$, $f_{0}^{B\to\pi}(q^{2})$ and
$f_{T}^{B\to\pi}(q^{2})$ (conventionally called vector, scalar and
tensor form factors, respectively), of which, the first two and
$f_{T}^{B\to\pi}(q^{2})$ parameterize, respectively, the matrix
elements of the vector and the tensor currents as
\begin{eqnarray}\label{eq:MatrixElements}
&&\langle \pi(p)|\bar{d}\gamma_{\mu}b|B(p+q)\rangle=(2p+q)_{\mu}f_{+}^{B\to\pi}(q^{2})+\frac{m_{B}^{2}-m_{\pi}^{2}}{q^{2}}q_{\mu}\left(f_{0}^{B\to\pi}(q^{2})-f_{+}^{B\to\pi}(q^{2})\right),\nonumber\\
&&\langle
\pi(p)|\bar{d}\sigma_{\mu\nu}q^{\nu}b|B(p+q)\rangle=i\left((2p+q)_{\mu}q^{2}-\left(m_{B}^{2}-m_{\pi}^{2}\right)q_{\mu}\right)\frac{f_{T}^{B\to\pi}(q^{2})}{m_{B}+m_{\pi}},
\end{eqnarray}
where the 4-momentum assignment is specified in brackets, and
$m_{B}$ ($m_{\pi}$) denotes the $B$ ($\pi)$ meson mass. Leaving
aside potential uncertainties with the existing QCD approaches to
form factors for heavy-to-light $B$ decays, the key problem is that
none of them is applicable in the entire $q^2$ region. Whereas
lattice QCD (LQCD) simulation, as a rigorous approach, could make
prediction at large $q^2$, QCD LCSR \cite{Balitsky,
LSCR-V.L.Chernyak} and perturbative QCD (pQCD)
\cite{H.N.Li-1201.5066} approaches are applicable for low and
intermediate $q^2$. In \cite{BPill-J.J.Wang-0711.0321} the LCSR
computations presented by \cite{P.Ball-0406232} were extrapolated to
the high $q^2$ region by using the $B^*$-dominance assumption, to
make an estimate for $B\to\pi\ell^{+}\ell^{-}$. The same was done in
pQCD approach \cite{BPill-Z.J.Xiao-1207.0265}. To enhance prediction
accuracy for the $B\to \pi$ form factors in the whole physical
region, a quasi-model-independent approach
\cite{BPill-A.Ali-1312.2523} has recently been suggested, in which
use was made of available experimental measurements as well as
theoretical predictions from LQCD simulation and other scenarios.
For example, the shape of the vector form factor was extracted from
the experimental data on the $B\to\pi\ell\nu_{\ell}$ semileptonic
decays; for determination of $q^2$-behavior of the tensor form
factor, the constraints were utilized from the LQCD data on the
corresponding $B\to K$ form factor combined with an ansatz on
$SU_F(3)$ symmetry breaking, and the heavy quark symmetry in the
large recoil limit. However, the vector form factor obtained therein
by data-fitting is based on use of the result from the CKM unitarity
fits \cite{PDG2012}, $|V_{ub}|=(3.51^{+0.15}_{-0.14})\times10^{-3}$,
which is incompatible with inclusive determinations. To say at
least, even if it reflects the true value of $|V_{ub}|$, the form
factor shape extracted experimentally requires a dynamical
interpretation.

Given the fact that LCSR approach has exhibited a stronger
predictive power in its applications to numerous exclusive
processes, and could be substantially complementary to LQCD
simulation in the aspect of predicting the form factors, in this
study we attempt to combine the LCSR calculations with the available
LQCD results and the analyticity of the form factors, to revaluate
$f_{+}^{B\to\pi}(q^2)$, $f_{T}^{B\to\pi}(q^2)$ and
$f_{0}^{B\to\pi}(q^2)$, and then in naive factorization explore the
$B\to \pi\ell^+\ell^-$ decays and make comparison with the recent
study of \cite{BPill-A.Ali-1312.2523}.

Calculations of the form factors in question have already been
undertaken many times within the LCSR framework. One can be referred
to \cite{LSCR-Z.H.Li-1206.0091, Bharucha} for a recent application
of this approach. At QCD next-leading-order (NLO) level for twist-2
and -3, the first complete study on these form factors was put
forward in \cite{P.Ball-0406232}, with the pole mass for the
underlying heavy quark in light-cone operator product expansion
(OPE) calculation of the correlation functions. To the same
accuracy, instead using the $\overline{\mathrm{MS}}$ mass the
authors of \cite{LSCR-A.Khodja-0801.1796} furnished an updated
computation, with which $|V_{ub}|$ was extracted from the the BarBar
data \cite{LSCR-A.Khodja-1103.2655}. Here we would like to take an
alternative version suggested in \cite{LSCR-V.L.Chernyak,
LCSR-Z.H.Li, LSCR-Z.H.Li-1206.0091}, in which a certain chiral
current correlator is so chosen that the twist-3 and -5 components
of the pionic light-cone distribution amplitudes (DAs) do not
contribute and thus the resulting sum rules receive less pollution
than in the case of the standard correlation functions. It should be
added that the calculations with LCSR involve soft-overlap as well
as hard-exchange components, and the former plays a predominant
role. It forms a striking contrast to the situation when applying
pQCD approach \cite{H.N.Li-1201.5066}, in which hard-exchange
dominates.

This paper is organized as follows. The following section
encompasses a concise derivation of the LCSRs for the $B\to \pi$
vector, scalar and tensor form factors and numerical analysis. In
section 3 we turn to the discussion about the shapes of the form
factors in the whole kinematically accessible region. In section 4,
we apply our findings to estimate the decay rates and branching
ratios for the $B\to\pi$ dileptonic decays, including the ditau
modes, and also the partial branching ratios in some chosen $q^2$
bins. The final section is devoted to a concluding remark.
\section{LCSR calculation of the $B\to\pi\ell^+\ell^-$ form factors}
Essentially, LCSR approach is through the twist expansion of, say, a
vacuum-to-pion correlation function in the small light-cone distance
$x^2\approx0$ and in the strong coupling $\alpha_s$, which works
effectively out some of the problems with the short distance
($x\approx 0$) expansion in terms of vacuum condensates. To validate
the light-cone OPE, the higher-twist terms are required to be
suppressed. In the heavy quark expansion \cite{P.Ball-0608116}, it
is seen readily that higher-twist contributions increase with $q^2$
so that for larger $q^2$ the twist hierarchy breaks down. The
accessible kinematical region can be approximately fixed at $0\leq
q^2\leq 12-14~\mathrm{GeV}^2$. In the ensuing LCSR calculation, we
will restrict ourself to the interval $0\leq q^2\leq
12~\mathrm{GeV}^2$, to ensure the validity of results.
\subsection{NLO QCD calculation}
We employ the following vacuum-to-pion correlation functions to
achieve a LCSR estimate of the $B\to \pi \ell^+\ell^-$ form factors,
\begin{eqnarray}\label{eq:CorrFuncF+0}
F_{\mu}(p,q)&=&i\int d^{4}xe^{iq\cdot x}\langle \pi(p)|T\{\bar{d}(x)\gamma_{\mu}(1+\gamma_{5})b(x),m_{b}\bar{b}(0)i(1+\gamma_{5})u(0)\}|0\rangle\nonumber\\
&=&F\left(q^{2},(p+q)^{2}\right)p_{\mu}+\overline{F}\left(q^{2},(p+q)^{2}\right)q_{\mu},\\\label{eq:CorrFuncFT}
\widetilde{F}_{\mu}(p,q)&=&i\int d^{4}xe^{iq\cdot x}\langle \pi(p)|T\{\bar{d}(x)i\sigma_{\mu\nu}q^{\nu}(1+\gamma_{5})b(x),m_{b}\bar{b}(0)i(1-\gamma_{5})u(0)\}|0\rangle\nonumber\\
&=&\widetilde{F}\left(q^{2},(p+q)^{2}\right)\left[q_{\mu}(q\cdot
p)-p_{\mu}q^{2}\right],
\end{eqnarray}
with $m_b$ being the $b$ quark mass, and take the chiral limit
$m_{\pi}=0$ for the pion mass throughout the derivation. Note that a
T-product of chiral currents, which keeps the hadronic contribution
to the correlation function positive definite, is substituted for
the corresponding one adopted in the standard approach. In a large
space-like momentum region $(p+q)^{2}\ll0 $ and the effective
$q^{2}$ interval, the correlation functions can be expanded in the
small light-cone distance $x^2\approx 0$, and however, the operator
replacements result in an explicitly different OPE, in which,
especially, no twist-3 and -5 component is involved, as
aforementioned and seen below. As a result, the invariant functions
$F(q^{2},(p+q)^{2})$, $\overline{F}(q^{2},(p+q)^{2})$ and
$\widetilde{F}(q^{2},(p+q)^{2})$, which have the generic expansion
in $\alpha_s$,
\begin{eqnarray}
H^{QCD}\left(q^2,(p+q)^2\right)=H_{0}^{QCD}\left(q^{2},
(p+q)^2\right)+\frac{\alpha_{s}C_{F}}{4\pi}H_{1}^{QCD}\left(q^{2},
(p+q)^2\right)+\cdot\cdot\cdot.
\end{eqnarray}
are made accessible at twist-5 level with the existing findings of
the twist-2 and -4 DAs. The resulting difference in hadronic
expression is that there are the additional terms due to the
complete set of scalar $(0^+)$ $B$ meson states. However, this
causes no problem, because they are located far away from the lowest
pseudoscalar state and therefore their contributions can safely be
absorbed in a dispersion integral.

According to the standard procedure of sum rule calculation, the
form factors in question are accessible. Assuming the quark-hadron
duality and matching the OPE form of the correlation function \eqref{eq:CorrFuncF+0}
\begin{eqnarray}
F_{\mu}^{QCD}(p,q)=F^{QCD}\left(q^{2},(p+q)^{2}\right)p_{\mu}+\overline{F}^{QCD}\left(q^{2},(p+q)^{2}\right)q_{\mu},
\end{eqnarray}
with the corresponding hadronic one, which follows from inserting
the complete sets of both the pseudoscalar and scalar states between
the currents of \eqref{eq:CorrFuncF+0} and then isolating
the pole contribution from the lowest $B$ meson, we obtain
\begin{eqnarray}
\frac{2m_{B}^{2}f_{B}f^{B\to
\pi}_+(q^{2})}{m_{B}^{2}-(p+q)^{2}}=\frac{1}{\pi}\int_{m_b^2}^{s_0}\frac{\mathrm{Im}F^{QCD}(q^2,s)}{s-(p+q)^2}ds,
\end{eqnarray}
\begin{eqnarray}
\frac{m_{B}^{4}f_{B}}{q^2(m_B^2-(p+q)^2)}\left(f_{0}^{B\to\pi}(q^2)-\frac{m_B^2-q^2}{m_B^2}f_{+}^{B\to\pi}(q^2)\right)=\frac{1}{\pi}\int_{m_b^2}^{s_0}\frac{\mathrm{Im}\overline{F}^{QCD}(q^2,s)}{s-(p+q)^2}ds,
\end{eqnarray}
with $f_{B}$ being the decay constant defined as
$m_{B}^{2}f_{B}=\langle B|m_{b}\bar{b}i\gamma_{5}u|0\rangle$, $s_0$
an effective threshold to be determined and
$f_{0}^{B\to\pi}(0)=f_{+}^{B\to\pi}(0)$. On the Borel transformation
$(p+q)^{2}\rightarrow M^{2}$ for the above equations, we have the
sum rules for the vector and the scalar form factors:
\begin{eqnarray}\label{eq:f+Bpi}
f_{+}^{B\to\pi}(q^{2})=\frac{1}{2
m_{B}^{2}f_{B}}e^{m_{B}^{2}/M^{2}}F(q^2, M^2, s_0),
\end{eqnarray}
\begin{eqnarray}\label{eq:f0Bpi}
f_{0}^{B\to\pi}(q^2)=\frac{m_B^2-q^2}{m_B^2}f_{+}^{B\to\pi}(q^2)+\frac{q^2}{
m_B^4f_B}e^{m_{B}^{2}/M^{2}}\overline{F}(q^2, M^2, s_0),
\end{eqnarray}
where the functions $F(q^2, M^2, s_0)$ and $\overline{F}(q^2, M^2,
s_0)$ have the following form,
\begin{eqnarray}\label{eq:H-H0H1}
H(q^2, M^2, s_0)
&=&\frac{1}{\pi}\int_{m_b^2}^{s_0}ds~e^{-s/M^2}\mathrm{Im}H^{QCD}(q^2,s)\nonumber\\
&=&H_{0}
(q^{2},M^{2},s_0)+\frac{\alpha_{s}C_{F}}{4\pi}H_{1}(q^{2},M^{2},s_0)+\cdot\cdot\cdot.
\end{eqnarray}
For the correlation function \eqref{eq:CorrFuncFT}, a similar manipulation results in
the sum rule for the tensor form factor,
\begin{eqnarray}\label{eq:fTBpi}
f_{T}^{B\to\pi}(q^{2})=\frac{1}{2m_{B}f_{B}}e^{m_{B}^{2}/M^{2}}\widetilde{F}(q^2, M^2,
s_0),
\end{eqnarray}
with $\widetilde{F}(q^2, M^2, s_0)$ being defined the same as
$F(q^2, M^2, s_0)$ and $\overline{F}(q^2, M^2, s_0)$.

We are to do the OPE calculation at one-loop level for twist-2. The
LO functions, $F_{0} (q^{2},M^{2},s_0)$,
$\overline{F}_{0}(q^{2},M^{2},s_0)$ and
$\widetilde{F}_{0}(q^{2},M^{2},s_0)$, are easy to get, by
contracting the $b$ quark fields of (2) and (3) to the free quark
propagator plus a correction term from one-gluon emission and using
the definition of the pion DAs \cite{P.Ball-0502115, P.Ball-0507076,
LSCR-DK-Khodjamirian-0907.2842}. Then it becomes clear that the
twist-3 and-5 contributions vanish due to the Dirac structures. The
results read,
\begin{eqnarray}\label{eq:f+(0)}
F_{0}(q^{2},M^{2},s_0)&=& 2m_{b}^{2}f_{\pi}\int_{u_{0}}^{1}du
e^{-\frac{m_{b}^{2}-q^{2}\bar{u}}{uM^{2}}}\left\{\frac{\varphi_{\pi}(u)}{u}+\frac{1}{m_{b}^{2}-q^{2}}\left(-\frac{m_{b}^{2}u}{4(m_{b}^{2}-q^{2})}\frac{d^{2}\phi_{4\pi}(u)}{du^{2}}\right.\right.\nonumber\\
&+&\left.\left.u\psi_{4\pi}(u)+\int_{0}^{u}dv
\psi_{4\pi}(v)-\frac{d}{du}J_{4\pi}(u)\right)\right\},\\\label{eq:f0(0)}
\overline{F}_{0}(q^{2},M^{2},s_0)&=&
2m_{b}^{2}f_{\pi}\int_{u_{0}}^{1}du
e^{-\frac{m_{b}^{2}-q^{2}\bar{u}}{uM^{2}}}\frac{1}{m_{b}^{2}-q^{2}}\psi_{4\pi}(u),\\\label{eq:fT(0)}
\widetilde{F}_{0}(q^{2},M^{2},s_0)&=&
2m_{b}f_{\pi}\int_{u_{0}}^{1}du
e^{-\frac{m_{b}^{2}-q^{2}\bar{u}}{uM^{2}}}\left\{\frac{\varphi_{\pi}(u)}{u}+\frac{1}{m_{b}^{2}-q^{2}}\left(\frac{1}{4}\frac{d\phi_{4\pi}(u)}{du}\right.\right.\nonumber\\
&-&\left.\left.\frac{m_{b}^{2}u}{2(m_{b}^{2}-q^{2})}\frac{d^{2}\phi_{4\pi}(u)}{du^{2}}-\frac{d}{du}\widetilde{J}_{4\pi}(u)\right)\right\}.
\end{eqnarray}
In the above, $\bar{u}=1-u$, $u_{0}=(m_{b}^{2}-q^{2})/(s_0-q^{2})$,
$f_{\pi}$ indicates the pionic decay constant, $J_{4\pi}(u)$ and
$\widetilde{J}_{4\pi}(u)$ are two integral functions:
\begin{eqnarray}\label{eq:FT}
J_{4\pi}(u)&=&\int_{0}^{u}d\alpha_{1}\int_{\frac{u-\alpha_{1}}{1-\alpha_{1}}}^{1}\frac{dv}{v}
\left[2\Psi_{4\pi}(\alpha_{i})+2\tilde{\Psi}_{4\pi}(\alpha_{i})\right.\nonumber\\
&&\left.\left.-\Phi_{4\pi}(\alpha_{i})-\tilde{\Phi}_{4\pi}(\alpha_{i})\right]\right|_{\begin{subarray}{l}
 \alpha_{2}=1-\alpha_{1}-\alpha_{3}\\
 \alpha_{3}=(u-\alpha_{1})/v\end{subarray}},\\\label{eq:FT}
\widetilde{J}_{4\pi}(u)&=&\int_{0}^{u}d\alpha_{1}\int_{\frac{u-\alpha_{1}}{1-\alpha_{1}}}^{1}\frac{dv}{v}
\left[2\Psi_{4\pi}(\alpha_{i})+2(1-2v)\tilde{\Psi}_{4\pi}(\alpha_{i})\right.\nonumber\\
&&\left.\left.-(1-2v)\Phi_{4\pi}(\alpha_{i})-\tilde{\Phi}_{4\pi}(\alpha_{i})\right]\right|_{\begin{subarray}{l}
 \alpha_{2}=1-\alpha_{1}-\alpha_{3}\\
 \alpha_{3}=(u-\alpha_{1})/v\end{subarray}};
\end{eqnarray}
$\varphi_{\pi}(u)$ denotes the leading twist-2 DA, while
$\phi_{4\pi}(u)$, $\psi_{4\pi}(u)$ and those functions included in
the integrands of (15) and (16) have all twist-4.

By substituting \eqref{eq:f+(0)}, \eqref{eq:f0(0)} and
\eqref{eq:fT(0)}, respectively, into \eqref{eq:f+Bpi},
\eqref{eq:f0Bpi} and \eqref{eq:fTBpi}, the resulting sum rules for
$f_{+}^{B\to\pi}(q^2)$, $f_{0}^{B\to\pi}(q^2)$ and
$f_{T}^{B\to\pi}(q^2)$ respect, up to the higher twist and QCD
radiative corrections, the following relations which are similar to
the observations in the limits of heavy quark mass and large recoil
energy \cite{Charles-014001, C.W.Bauer-0011336}:
\begin{eqnarray}
f_{0}^{B\to\pi}(q^2)&=&\frac{m_B^2-q^2}{m_B^2}f_{+}^{B\to\pi}(q^2),\\
f_{T}^{B\to\pi}(q^2)&=&\frac{m_B}{m_b}f_{+}^{B\to \pi}(q^2)\nonumber\\
&=&\frac{m_B^3}{m_b}\left(\frac{f_{+}^{B\to\pi}(q^2)-f_{0}^{B\to\pi}(q^2)}{q^2}\right),
\end{eqnarray}
so that in such an approximation only one independent form factor is
necessary for describing the non-perturbative QCD dynamics involved
in the $B\to \pi$ transitions. In effect, in the case of using the
standard correlation functions the same relations hold numerically
approximately, despite not explicitly appearing. All these provide
an important validity check of the present approach.

\begin{figure}
\includegraphics[scale=0.9]{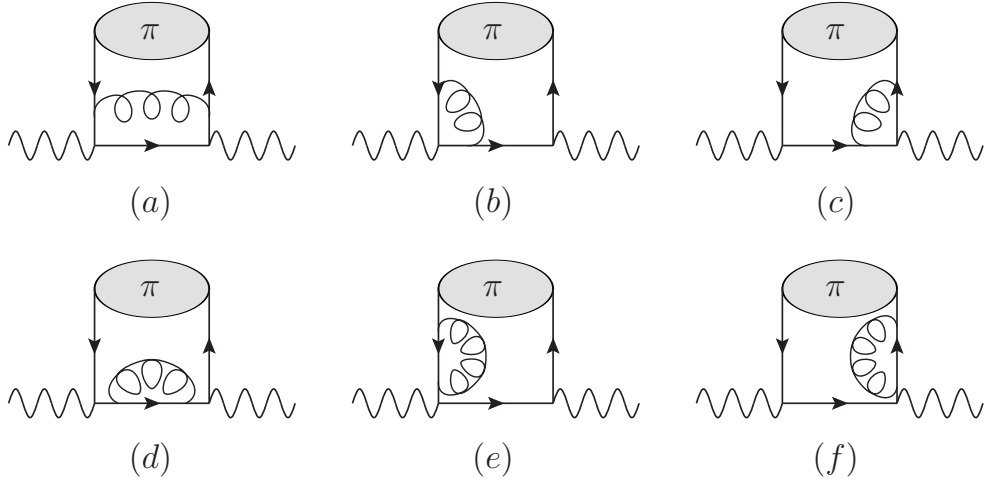}
\caption{One-loop Feynman diagrams contributing to the correction
functions.}\label{fig:feynman}
\end{figure}

For getting the NLO corrections, $F_1(q^2, M^2, s_0)$,
$\overline{F}_1(q^2, M^2, s_0)$ and $\widetilde{F}_1(q^2, M^2,s_0)$,
we turn to calculation of the invariant functions
$F_1^{QCD}(q^2,(p+q)^2)$, $\overline{F}_1^{QCD}(q^2,(p+q)^2)$ and
$\widetilde{F}_1^{QCD}(q^2,(p+q)^2)$ (with the relevant Feynman
diagrams plotted in Fig.\ref{fig:feynman}). Apparently they can be
expressed uniformly as a convolution of the corresponding hard
scattering amplitudes with the twist-2 DA: for example,
\begin{eqnarray}\label{eq:F(1)QCD}
F_{1}^{QCD}(q^{2},(p+q)^{2})=-f_{\pi}\int_{0}^{1}du
~T^H_{1}\left(q^{2},(p+q)^{2},u\right)\varphi_{\pi}(u),
\end{eqnarray}
and hence our task boils down to computing the scattering functions
$T^H_{1}(q^{2},(p+q)^{2},u)$,
$\overline{T}^H_{1}(q^{2},(p+q)^{2},u)$ and
$\widetilde{T}^H_{1}(q^{2},(p+q)^{2},u)$.

To this end, we take the Feynman gauge, and adopt the dimensional
regularization and the $\overline{\mathrm{MS}}$ mass
$\overline{m}_b$ for the underlying $b$ quark. In fact, the same
prescription has been employed to investigate the QCD radiative
correction to the vector form factor at $q^2=0$ in
\cite{LSCR-Z.H.Li-1206.0091}, where a detailed derivation of
$F_{1}^{QCD}(q^{2}=0,(p+q)^{2})$ is presented. Using the technic
described therein, the invariant functions in question could be
worked out. In what follows, we just highlight the key points and
main results in the NLO calculation.

At first we concentrate on a discussion of
$F_{1}^{QCD}(q^{2},(p+q)^{2})$. A straightforward calculation shows
that there are both ultraviolet (UV) and infrared (IR) divergences
to deal with in $T^H_{1}(q^{2},(p+q)^{2},u)$. By performing the mass
renormalization, $m_b\to Z_m \overline{m}_b$ with $Z_m$ being the
familiar renormalization constant, the UV divergence, as expected,
is precisely offset by the one appearing in the resulting LO term,
leading to an UV finite hard scattering amplitude written down in
terms of the $\overline{\mathrm{MS}}$ mass that we denote by $m_b$
hereafter, unless otherwise stated. As for the IR divergence term,
it can be eliminated by replacing the bare quantity
$\varphi_{\pi}(u)$ of \eqref{eq:F(1)QCD} with a renormalized DA
$\varphi_{\pi}(u,\mu)$. As a result, we are left with the invariant
function $F_{1}^{QCD}(q^{2},(p+q)^{2})$ expressed by the convolution
of $\varphi_{\pi}(u,\mu)$ with a scale-dependent NLO hard kernel
$T^H_{1}(\eta_{1},\eta_{2},u,\mu)$ obtained as following,
\begin{eqnarray}
T^H_{1}(\eta_{1},\eta_{2},u,\mu)&=&4\left(\frac{1}{1-\eta}+\frac{\eta_{2}-1}{u(\eta_{2}-\eta_{1})^{2}}\right)L(\eta_{1})+4\left(\frac{1}{1-\eta}-\frac{1-\eta_{1}}{\bar{u}(\eta_{2}-\eta_{1})^{2}}\right)
L(\eta_{2})\nonumber\\
&-&4\left(\frac{2}{1-\eta}+\frac{\eta_2-1+u(\eta_1-\eta_2)}{u\bar{u}(\eta_{1}-\eta_{2})^{2}}\right)L(\eta)\nonumber\\
&-&\frac{4}{\eta_{2}}\left(\frac{\eta_{2}-1}{\eta-1}-\frac{\eta_{2}-1}{\bar{u}(\eta_{2}-\eta_{1})}\right)\mathrm{ln}(1-\eta_{2})
-\frac{2}{\eta_{2}}\left(\frac{\eta_{2}-2}{\eta}-\frac{\eta_{2}}{\eta^{2}}+\frac{2(\eta_{2}-1)}{\bar{u}(\eta_{2}-\eta_{1})}\right)\nonumber\\
&\times&\mathrm{ln}(1-\eta)-\frac{2(\eta+1)}{(\eta-1)^{2}}\left(3\mathrm{ln}\frac{m_{b}^{2}}{\mu^{2}}-\frac{3\eta+1}{\eta}\right),
\end{eqnarray}
where $\eta_{1}=q^{2}/{m_{b}^{2}}$, $\eta_{2}=(p+q)^{2}/m_{b}^{2}$,
$\eta=\eta_{1}+u(\eta_{2}-\eta_{1})$, and $L(x)$ indicates a linear
combination of the form:
\begin{eqnarray}
L(x)=\mathrm{Li}_{2}(x)+\mathrm{ln}^{2}(1-x)+\mathrm{ln}(1-x)\left(\mathrm{ln}\frac{m_{b}^{2}}{\mu^{2}}-1\right)\nonumber,
\end{eqnarray}
with the dilogarithm
$\mathrm{Li}_2(x)=-\int_0^x~dx~\frac{\mathrm{ln}(1-x)}{x}$. It
should be understood that here and hereafter the factorization scale
is specified to be equal to the renormalization one. Putting
everything together yields the desired result,
\begin{eqnarray}
F_1(q^2,M^2,s_0)=-\frac{f_{\pi}}{\pi}\int_{m_{b}^{2}}^{s_0}ds
e^{-s/M^{2}}\int_{0}^{1} du\mathrm{Im}T_{1}^H(\eta_{1},
\eta_{2},u,\mu)\varphi_{\pi}(u,\mu),
\end{eqnarray}
with $\eta_{2}=s/m_b^2>1$ (in the following the same should be
understood when taking imaginary part for a hard kernel), and
\begin{eqnarray}\label{eq:ImT+}
&&\frac{1}{2\pi}\mathrm{Im}T^H_{1}(\eta_{1},\eta_{2},u,\mu)=\delta(1-\eta)\left[6-3\mathrm{ln}\frac{m_{b}^{2}}{\mu^{2}}-\frac{7}{3}\pi^{2}
+2\mathrm{Li}_{2}(\eta_{1})-2\mathrm{Li}_{2}(1-\eta_{2})\nonumber\right.\\
&&~~~~~~~~+2\left(\mathrm{ln}^{2}(1-\eta_{1})+\mathrm{ln}^{2}(\eta_{2}-1)\right)-2\left(\mathrm{ln}{\eta_{2}}+\frac{1-\eta_{2}}{\eta_{2}}\right)
\mathrm{ln}(\eta_{2}-1)\nonumber\\
&&~~~~~~~~-\left.2\mathrm{ln}\left((1-\eta_{1})(\eta_{2}-1)\right)\left(1-\mathrm{ln}\frac{m_{b}^{2}}{\mu^{2}}\right)-2
\left(4-3\mathrm{ln}\frac{m_{b}^{2}}{\mu^{2}}\right)\left(1+\frac{\mathrm{d}}{\mathrm{d}\eta}\right)\right]\nonumber\\
&&~~~~~~~~+\theta(\eta-1)\left[\left.\frac{4}{\eta-1}\right|_{+}\left(\mathrm{ln}\left(\frac{(\eta-1)^{2}}{\eta}\frac{m_{b}^{2}}{\mu^{2}}\right)-1\right)-\left.\frac{2}{\eta-1}\right|_{+}\left(\mathrm{ln}\left(\frac{(\eta_{2}-1)^{2}}{\eta_{2}}\frac{m_{b}^{2}}{\mu^{2}}\right)-\frac{1}{\eta_{2}}\right)\right.\nonumber\\
&&~~~~~~~~-2\frac{1-\eta_{1}}{(\eta_{2}-\eta_{1})(\eta_{2}-\eta)}\left(\mathrm{ln}\left(\frac{(\eta_{2}-1)^{2}}{\eta_{2}}\frac{m_{b}^{2}}{\mu^{2}}\right)-1\right)+\frac{1}{\eta}\left(\frac{1}{\eta}+\frac{2}{\eta_{2}}-1\right)\nonumber\\
&&~~~~~~~~+\left.2\frac{1+\eta-\eta_{1}-\eta_{2}}{(\eta_{1}-\eta)(\eta_{2}-\eta)}\left(\mathrm{ln}\left(\frac{(\eta-1)^{2}}{\eta}\frac{m_{b}^{2}}{\mu^{2}}\right)-1\right)
\right]\nonumber\\
&&~~~~~~~~+\theta(1-\eta)
\left[2\left.\left(\mathrm{ln}\frac{\eta_{2}}{(\eta_{2}-1)^{2}}+\frac{1}{\eta_{2}}-\mathrm{ln}\frac{m_{b}^{2}}{\mu^{2}}\right)\frac{1}{\eta-1}\right|_{+}\right.\nonumber\\
&&~~~~~~~~-\left.2\frac{1-\eta_{1}}{(\eta_{1}-\eta_{2})(\eta_{2}-\eta)}\left(\mathrm{ln}\frac{\eta_{2}}{(\eta_{2}-1)^{2}}+1-\mathrm{ln}\frac{m_{b}^{2}}{\mu^{2}}\right)
-2\frac{1}{\eta_{2}-\eta}\frac{1-\eta_{2}}{\eta_{2}}\right].
\end{eqnarray}
Note the operation,
\begin{eqnarray}
\left.\frac{F(\eta)}{1-\eta}\right|_+=\frac{F(\eta)-F(1)}{1-\eta},
\end{eqnarray}
is introduced to avert the redundant IR divergences generated by
taking the imaginary part. Finally, using the known
$F_{1}(q^{2},M^2,s_0)$ and $F_{0}(q^{2},M^2,s_0)$ we achieve the
function $F(q^{2},M^2,s_0)$ with $\mathcal{O}(\alpha_s)$ accuracy,
where the changes with scale compensate each other of the hard
kernel and the twist-2 DA, having QCD factorization observed.

Contrasted with the above situation, neither UV nor IR divergences
appear in the calculation of $\overline{F}_{1}^{QCD}(q^{2},
(p+q)^{2})$. The NLO hard kernel reads as,
\begin{eqnarray}
\overline{T}_{1}^H(\eta_{1},\eta_{2},u)
&=&2\left[\frac{\eta_{1}^{2}-\eta_{1}\eta_{2}-(1-\eta_{1})(\eta_{2}-\eta_{1})\mathrm{ln}(1-\eta_{1})}{\eta_{1}^{2}(1-\eta)}-
\frac{(1-\eta_{1})(\eta_{1}+\eta_{2})\mathrm{ln}(1-\eta_{1})}{u\eta_{1}^{2}(\eta_{2}-\eta_{1})}\right.\nonumber\\
&+&\left.2\frac{(\eta_{2}-1)\mathrm{ln}(1-\eta_{2})}{\bar{u}\eta_{2}(\eta_{2}-\eta_{1})}-\frac{(\eta-1)(\eta_{2}+\eta)\mathrm{ln}(1-\eta)}
{u\bar{u}(\eta_{2}-\eta_{1})\eta^{2}}-\frac{\eta_{2}-\eta_{1}}{\eta_{1}\eta}\right].
\end{eqnarray}
From this, we derive the NLO function $\overline{F}_1(q^2,M^2,s_0)$,
\begin{eqnarray}\label{eq:F01}
\overline{F}_1(q^2,M^2,s_0)=
-\frac{f_{\pi}}{\pi}\int_{m_{b}^{2}}^{s_0}ds
e^{-s/M^{2}}\int_{0}^{1} du\mathrm{Im}\overline{T}_{1}^H(\eta_{1},
\eta_{2},u)\varphi_{\pi}(u,\mu),
\end{eqnarray}
\begin{eqnarray}\label{eq:ImT0}
\frac{1}{2\pi}\mathrm{Im}\overline{T}_{1}^H(\eta_{1},\eta_{2},u)
&=&\delta(1-\eta)\left[1-\frac{\eta_{2}}{\eta_{1}}-\frac{(\eta_{1}-1)(\eta_{1}-\eta_{2})\mathrm{ln}(1-\eta_{1})}{\eta_{1}^{2}}\right]\nonumber\\
&+&\theta(\eta-1)\left[\frac{2(\eta_{2}-1)}{\eta_{2}(\eta_{2}-\eta)}-\frac{(\eta-1)(\eta_{2}+\eta)}{u\bar{u}\eta^{2}(\eta_{2}-\eta_{1})}\right]\nonumber\\
&+&\theta(1-\eta)\frac{2(\eta_{2}-1)}{\eta_{2}(\eta_{2}-\eta)}.
\end{eqnarray}
As a consequence, the complete function $\overline{F}(q^{2},M^{2},s_0)$
has a QCD factorized form, in which the hard kernel is scale-independent.

To proceed, we embark upon discussing the case of the tensor form
factor. In dealing with the hard amplitude
$\widetilde{T}_{1}^{H}(q^2,(p+q)^2,u)$, we find that in addition to
an IR divergence which can be removed as in the case of
${T}_{1}^{H}(q^2,(p+q)^2,u)$, there is an UV divergence left after
the mass renormalization. It is not surprising because the effective
weak operator of the related correlation function is a combination
of the tensor and the pseudo-tensor operators, which require a
renormalization, as compared with the vector (axial-vector)
operator. On taking this point into account, the divergence, indeed,
could be canceled out by that entering the renormalization constant
of the effective current, which, from another perspective, provides
a confirmation of the calculation. We have the hard kernel,
\begin{eqnarray}
&&\widetilde{T}_{1}^{H}(\eta_{1},\eta_{2},u,\mu)=
\frac{4}{m_b}\left[\left(\frac{1}{1-\eta}+\frac{\eta_{2}-1}{u(\eta_{2}-\eta_{1})^{2}}\right)L(\eta_{1})\right.\nonumber\\
&&~~~~~+\left(\frac{1}{1-\eta}-\frac{1-\eta_{1}}{\bar{u}(\eta_{2}-\eta_{1})^{2}}\right)L(\eta_{2})
+\left(\frac{1-\eta_{2}-u(\eta_{1}-\eta_{2})}{u\bar{u}(\eta_{1}-\eta_{2})^{2}}-\frac{2}{1-\eta}\right)L(\eta)\nonumber\\
&&~~~~~+\left(\frac{1-\eta_{1}}{u\eta_{1}(\eta_{2}-\eta_{1})}+\frac{1-\eta_{1}}{\eta_{1}(1-\eta)}\right)\mathrm{ln}(1-\eta_{1})
+\left(\frac{\eta_{2}-1}{\eta_{2}(1-\eta)}-\frac{\eta_{2}-1}{\bar{u}(\eta_{2}-\eta_{1})\eta_{2}}\right)\mathrm{ln}(1-\eta_{2})\nonumber\\
&&~~~~~-\left(\frac{1}{2\eta^{2}}-\frac{\eta_{2}-1}{\bar{u}(\eta_{2}-\eta_{1})\eta_{2}}+\frac{1-\eta_{1}}{u\eta_{1}(\eta_{2}-\eta_{1})}-\frac{2\eta_{2}+\eta_{2}\eta_{1}-2\eta_{1}}{2\eta_{1}\eta_{2}\eta}\right)\mathrm{ln}(1-\eta)\nonumber\\
&&\left.~~~~~-\left(\frac{1}{2(1-\eta)}+\frac{3}{(1-\eta)^{2}}\right)\mathrm{ln}\frac{m_{b}^{2}}{\mu^{2}}+\frac{1}{1-\eta}-\frac{1}{2\eta}+\frac{4}{(1-\eta)^{2}}\right],
\end{eqnarray}
and the imaginary part,
\begin{eqnarray}\label{eq:ImTT}
&&\frac{m_b}{4\pi}\mathrm{Im}\widetilde{T}_{1}^{H}(\eta_{1},\eta_{2},u,\mu)
=\delta(1-\eta)\left\{-\frac{7}{6}\pi^{2}+1+\mathrm{Li}_{2}(\eta_{1})-\mathrm{Li}_{2}(1-\eta_{2})
+\mathrm{ln}^{2}(1-\eta_{1})\right.\nonumber\\
&&~~~~~+\mathrm{ln}^{2}(\eta_{2}-1)-\mathrm{ln}\left((1-\eta_{1})(\eta_{2}-1)\right)\left(1-\mathrm{ln}\frac{m_{b}^{2}}{\mu^{2}}\right)+
\frac{1-\eta_{1}}{\eta_{1}}\mathrm{ln}(1-\eta_{1})\nonumber\\
&&\left.~~~~~-\left(\frac{1-\eta_{2}}{\eta_{2}}+\mathrm{ln}\eta_{2}\right)\mathrm{ln}(\eta_{2}-1)-\frac{1}{2}\mathrm{ln}\frac{m_{b}^{2}}{\mu^{2}}
+\left(-4+3\mathrm{ln}\frac{m_{b}^{2}}{\mu^{2}}\right)\frac{\mathrm{d}}{\mathrm{d}\eta}\right\}\nonumber\\
&&~~~~~+\theta(\eta-1)\left\{\frac{1-\eta_{1}}{(\eta_{2}-\eta_{1})(\eta-\eta_{2})}\left[\mathrm{ln}\left(\frac{(\eta_{2}-1)^{2}}{\eta_{2}}
\frac{m_{b}^{2}}{\mu^{2}}\right)-1\right]\right.\nonumber\\
&&~~~~~+\frac{1+\eta-\eta_{1}-\eta_{2}}{(\eta-\eta_{1})(\eta_{2}-\eta)}\left[\mathrm{ln}\left(\frac{(\eta-1)^{2}}{\eta}\frac{m_{b}^{2}}{\mu^{2}}\right)-1\right]
-\left[\mathrm{ln}\left(\frac{(\eta_{2}-1)^{2}}{\eta_{2}}
\frac{m_{b}^{2}}{\mu^{2}}\right)-\frac{1}{\eta_{2}}\right]\frac{1}{\eta-1}\Big|_{+} \nonumber\\
&&\left.~~~~~+2\left[\mathrm{ln}\left(\frac{(\eta-1)^{2}}{\eta}\frac{m_{b}^{2}}{\mu^{2}}\right)-1\right]\frac{1}{\eta-1}\Big|_{+}+\frac{\eta_{1}-1}
{\eta_{1}(\eta-\eta_{1})}-\frac{1}{2\eta^{2}}+\frac{\eta_{1}\eta_{2}+2(\eta_{2}-\eta_{1})}{2\eta
\eta_{1}\eta_{2}}\right\}\nonumber\\
&&~~~~~+\theta(1-\eta)\left\{\frac{1-\eta_{1}}{(\eta_{2}-\eta_{1})(\eta-\eta_{2})}\left[\mathrm{ln}\left(\frac{(\eta_{2}-1)^{2}}{\eta_{2}}
\frac{m_{b}^{2}}{\mu^{2}}\right)-1\right]\right.\nonumber\\
&&\left.~~~~~-\left[\mathrm{ln}\left(\frac{(\eta_{2}-1)^{2}}{\eta_{2}}
\frac{m_{b}^{2}}{\mu^{2}}\right)-\frac{1}{\eta_{2}}\right]\frac{1}{\eta-1}\Big|_{+}+\frac{1-\eta_{2}}{\eta_{2}(\eta_{2}-\eta)}\right\}.
\end{eqnarray}
Since the scale dependence of the perturbative kernel including the
NLO correction (27) does not cancel out that of
$\varphi_{\pi}(u,\mu)$, we achieve a scale-dependent factorization
form for $\widetilde{F}(q^2,M^2,s_0)$, with the NLO term,
\begin{eqnarray}
\widetilde{F}_1(q^2, M^2, s_0)=
-\frac{f_{\pi}}{\pi}\int_{m_{b}^{2}}^{s_0}ds
e^{-s/M^{2}}\int_{0}^{1} du\mathrm{Im}\widetilde{T}_{1}^H(\eta_{1},
\eta_{2},u,\mu)\varphi_{\pi}(u,\mu).
\end{eqnarray}

Completing our LCSR calculations of $f_{+}^{B\to\pi}(q^2)$,
$f_{0}^{B\to\pi}(q^2)$ and $f_{T}^{B\to\pi}(q^2)$, we associate with
(\ref{eq:f+Bpi}-\ref{eq:fTBpi}) the obtained LO expressions for
$F(q^2,M^2,s_0)$, $\overline{F}(q^2,M^2,s_0)$ and
$\widetilde{F}(q^2,M^2,s_0)$ and twist-2 NLO corrections.

Lastly, we make a few remarks: (1) Because of the vanishing
contribution of the subleading twist-3 components, the twist-4 terms
paly a subdominant role in the resulting sum rules. They are highly
suppressed by the factor of $1/(m_b^2-q^2)$ with respect to the
leading twist-2 ones, in small and intermediate kinematical region,
so that the LCSR expressions show a good twist hierarchy and thus
are well convergent. (2) Actually, no odd-twist component is
involved in the present approach to any order in
$\mathcal{O}(\alpha_s)$, as readily verified.
\subsection{Numerical discussion}
We proceed to do numerical analysis, starting with choice of input
parameters entering the LCSR expressions. Obviously, the leading
twist-2 DA, which obeys a conformal expansion in the Gegenbauer
polynomials as
\begin{eqnarray}
\varphi_{\pi}(u, \mu)=6u\bar{u}\left(1+\sum_{n=1}^{\infty}
a_{2n}^{\pi}(\mu)C_{2n}^{3/2}(u-\bar u)\right),
\end{eqnarray}
remains the most important source of uncertainty in the sum rule
computation. A keen interest is taken in the first two moment
parameters $a_2^{\pi}(\mu)$ and $a_4^{\pi}(\mu)$, since the
Gegenbauer polynomials of higher-degree, which are rapidly
oscillating, are usually considered less important from a
phenomenological point of view. Given that the existing
determinations from some nonperturbative approaches involve a large
uncertainty, one has attempted to acquire them by matching
theoretical computation of a physical observable, regardless of the
higher-moment corrections, with its experimental observation. For
example, fitting the LCSR for the pion electromagnetic form factor
to the experimental data obtains \cite{LSCR-A.Khodja-1103.2655}
$a_{2}^{\pi}(\mu=1\mathrm{GeV})=0.17\pm 0.08$ and
$a_{4}^{\pi}(\mu=1\mathrm{GeV})=0.06\pm 0.1$. Also, there exists
some effort in exploring the higher-moment effects
\cite{P.Ball-0502115, T.Huang-1303.2301, T.Huang-1305.7391,
X.G.Wu-0901.2636, S.S.Agaev-1012.4671}. A recent study of
$\gamma\gamma^{*}\to \pi^{0}$ form factor \cite{S.S.Agaev-1012.4671}
reveals that higher-moment terms do indeed play a minor role and
gives the three sets of fitted parameters at $\mu=1~\mathrm{GeV}$:
\begin{eqnarray}\label{3sets}
&(\mathrm{\uppercase\expandafter{\romannumeral1}})~&
a_{2}^{\pi}=0.130,a_{4}^{\pi}=0.244,
a_{6}^{\pi}=0.179,a_{8}^{\pi}=0.141,
a_{10}^{\pi}=0.116,a_{12}^{\pi}=0.099,\nonumber\\
&(\mathrm{\uppercase\expandafter{\romannumeral2}})~&
a_{2}^{\pi}=0.140,a_{4}^{\pi}=0.230,
a_{6}^{\pi}=0.180,a_{8}^{\pi}=0.050,\nonumber\\
&(\mathrm{\uppercase\expandafter{\romannumeral3}})~&
a_{2}^{\pi}=0.160,a_{4}^{\pi}=0.220, a_{6}^{\pi}=0.080.
\end{eqnarray}
Being aware that the fitted results given in the above and
\cite{LSCR-A.Khodja-1103.2655} stand just for the``effective" values
corresponding to different approximations to the Gegenbauer
expansion, and in a pQCD calculation of any hard exclusive process
involving a pion, the contribution of the $a_{2}^{\pi}(\mu)$ term
dominates the twist-2 part, we can have consistent estimates with
each other, while using these fitted DAs with and without
higher-moment terms to make prediction. We would like to employ as
input the parameter sets of \eqref{3sets}.

Concerning the twist-4 DAs, there are the following
parameterizations in terms of the two nonperturbative quantities
$\delta_{\pi}^{2}$ and $\varepsilon_{\pi}$:
\begin{eqnarray}
&&\phi_{4\pi}(u)=\frac{200}{3}\delta_{\pi}^{2}u^{2}\bar{u}^{2}+8\delta_{\pi}^{2}\varepsilon_{\pi}\left\{u\bar{u}(2+13u\bar{u})\right.\nonumber\\
&&~~~~~~~~~+\left.2u^{3}(10-15u+6u^{2})\mathrm{ln}u+2\bar{u}^{3}(10-15\bar{u}+6\bar{u}^{2})\mathrm{ln}\bar{u}\right\},\\
&&\psi_{4\pi}(u)=\frac{20}{3}\delta_{\pi}^{2}C_{2}^{\frac{1}{2}}(2u-1),\\
&&\Phi_{4\pi}(\alpha_{i})=120\delta_{\pi}^{2}\varepsilon_{\pi}(\alpha_{1}-\alpha_{2})\alpha_{1}\alpha_{2}\alpha_{3},\\
&&\Psi_{4\pi}(\alpha_{i})=30\delta_{\pi}^{2}(\mu)(\alpha_{1}-\alpha_{2})\alpha_{3}^{2}\left[\frac{1}{3}+2\varepsilon_{\pi}(1-2\alpha_{3})\right],\\
&&\tilde{\Phi}_{4\pi}(\alpha_{i})=-120\delta_{\pi}^{2}\alpha_{1}\alpha_{2}\alpha_{3}\left[\frac{1}{3}+\varepsilon_{\pi}(1-3\alpha_{3})\right],\\
&&\tilde{\Psi}_{4\pi}(\alpha_{i})=30\delta_{\pi}^{2}\alpha_{3}^{2}(1-\alpha_{3})\left[\frac{1}{3}+2\varepsilon_{\pi}(1-2\alpha_{3})\right].
\end{eqnarray}
We take the updated estimates \cite{P.Ball-1012.4671}
$\delta_{\pi}^{2}=(0.18\pm 0.06)~\mathrm{GeV^2}$ and
$\varepsilon_{\pi}=\frac{21}{8}\omega_{4\pi}~(\omega_{4\pi}=0.2\pm
0.1)$, normalized at $1~\mathrm{GeV}$.

The remaining parameters to need pinning down include the $b$ quark
mass and decay constant of $B$ meson. From a bottomonium sum rule
calculation at four-loop precision level \cite{J.H.Kuehn-0702103},
the yielded estimate,
$\overline{m}_b(\overline{m}_b)=4.164\pm0.025~\mathrm{GeV}$, is
extremely suitable as an input. As far as the latter goes, for
consistency and also for narrowing down the uncertainty due to that
quantity it is appropriate to take the two-point sum rule expression
in terms of the $b$ quark $\overline{\mathrm{MS}}$ mass and with
$\mathcal{O}(\alpha_s)$ accuracy, as given in
\cite{M.Jamin-0108135}. In addition, we employ \cite{PDG2014} the
measurement values, $f_{\pi}=130.41~\mathrm{MeV}$ and
$m_{B}=5.279~\mathrm{GeV}$, and two-loop running down from
$\alpha_s(M_z)=0.1185\pm 0.0006$ for the QCD coupling constant. The
factorization scale, in the light of the typical virtuality of the
underlying $b$ quark, is set at $\mu=3.0^{+1.5}_{-0.5}~\mathrm{GeV}$
in the case of both $f_{+}^{B\to\pi}(q^2)$ and
$f_{0}^{B\to\pi}(q^2)$, while for the scale dependent quantity
$f_{T}^{B\to\pi}(q^2)$, we estimate it at the scale
$\mu=4.8~\mathrm{GeV}$ for later convenience.

\begin{figure}[]
\subfigure[]{ \label{fig:f+MM}
\begin{minipage}[t]{0.5\textwidth}
\centering
\includegraphics[scale=0.6]{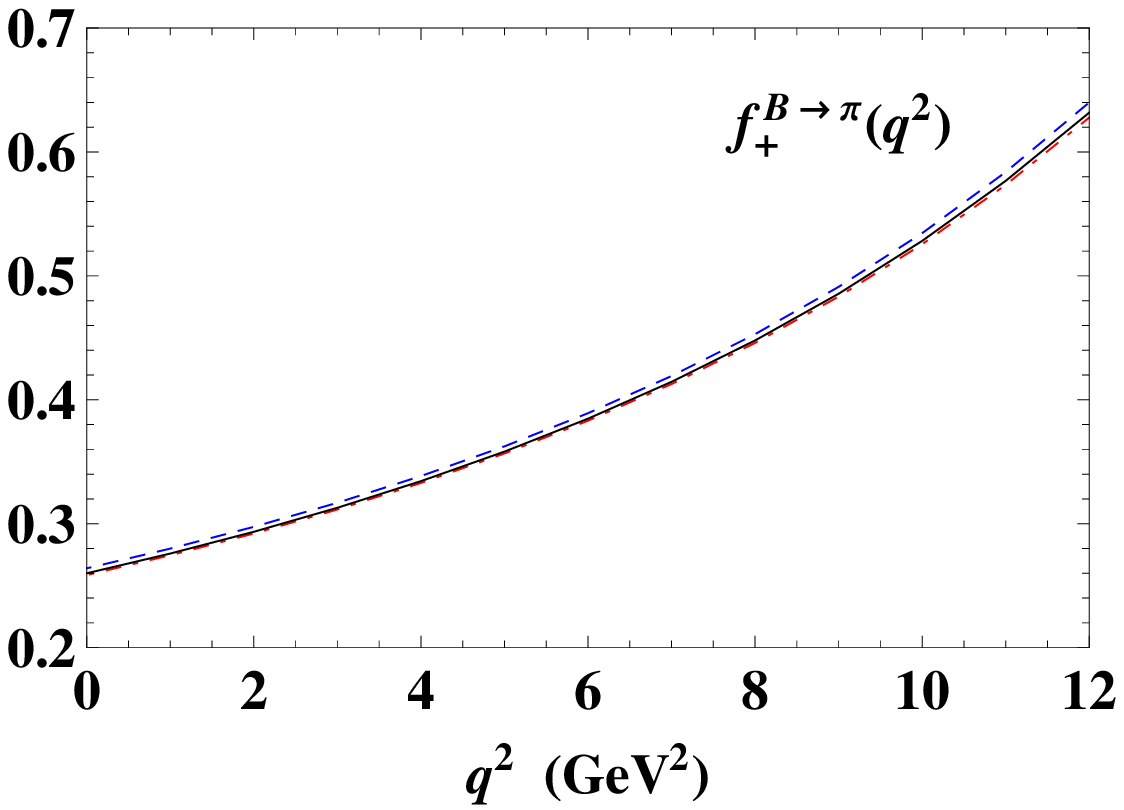}
\end{minipage}}%
\subfigure[]{ \label{fig:f0MM}
\begin{minipage}[t]{0.5\textwidth}
\centering
\includegraphics[scale=0.6]{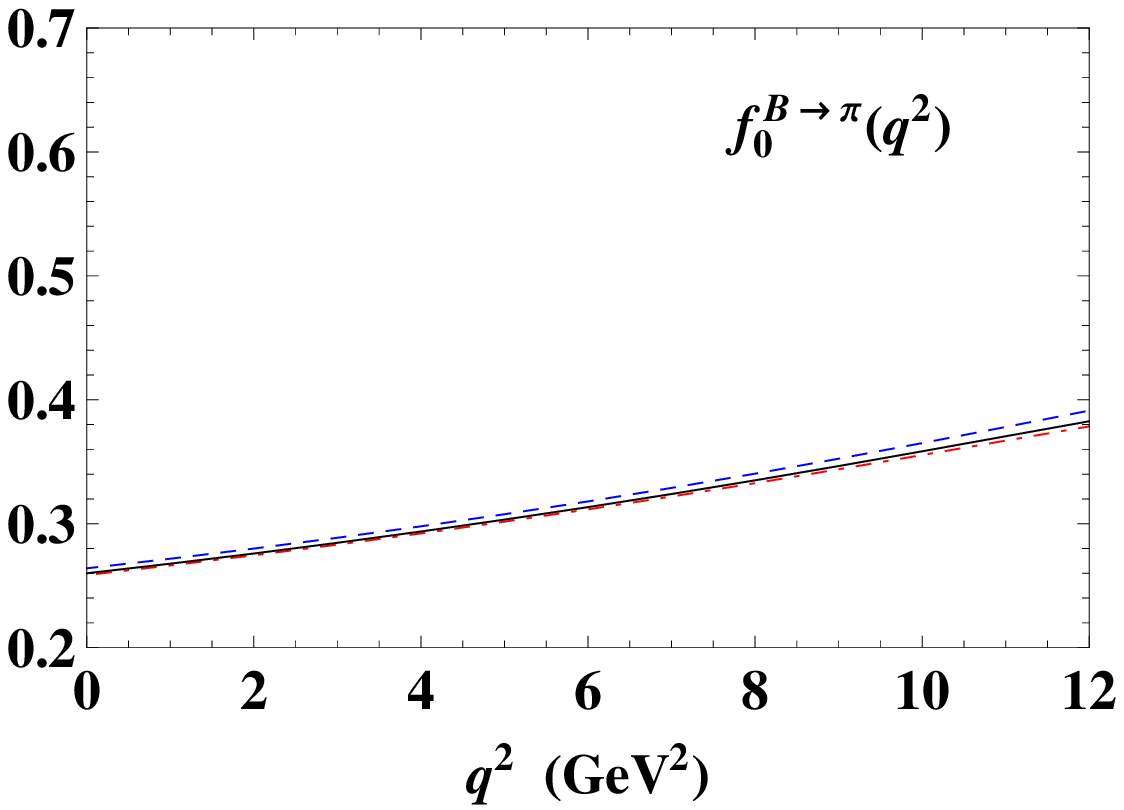}
\end{minipage}}%
\begin{center}
\subfigure[]{ \label{fig:fTMM}
\begin{minipage}[t]{0.5\textwidth}
\centering
\includegraphics[scale=0.6]{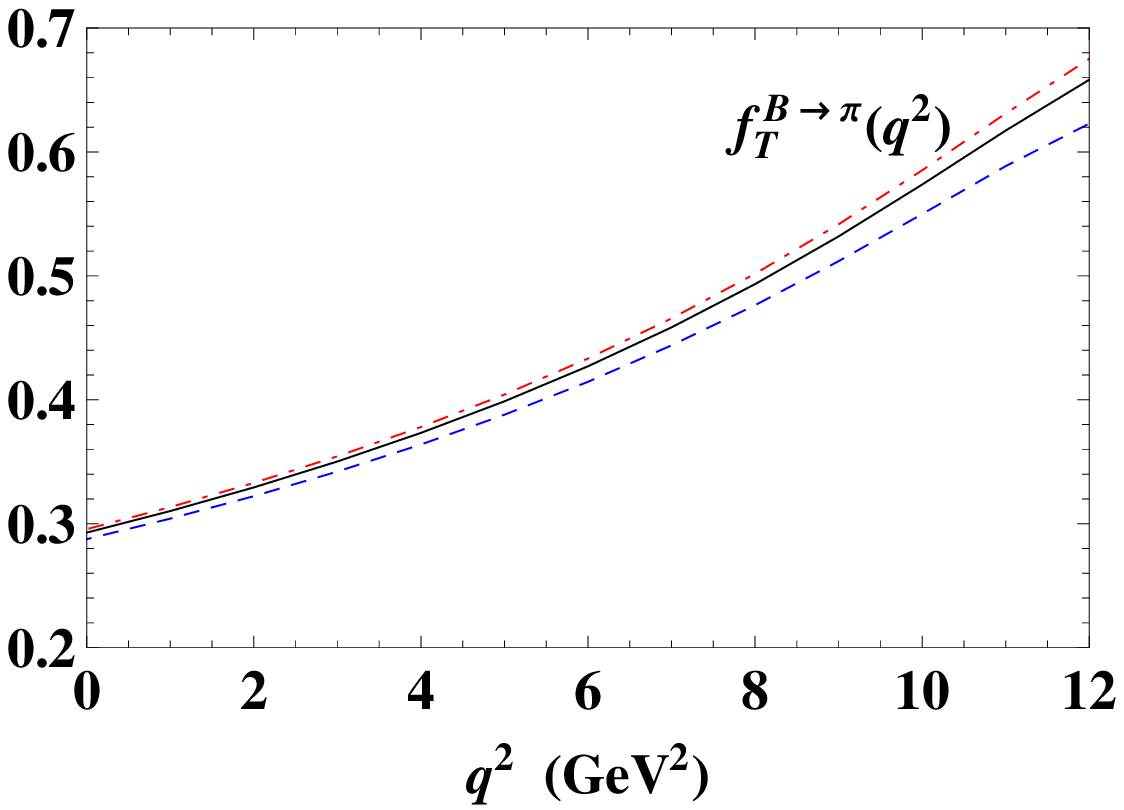}
\end{minipage}}%
\caption{Stability of the LCSRs for the $B\to \pi \ell^+\ell^-$ form
factors with respect to the variation of the Borel parameter $M^2$.
The solid lines indicate the central values and the regions between
the red dot-dashed and blue dashed lines do the
uncertainties.}\label{fig:f+0TMM}
\end{center}
\end{figure}
As two intrinsic parameters, the effective threshold $s_0$ and Borel
variable $M^2$ could be fixed in the standard procedure. Not being
an universal quantity, the threshold parameter has to be
independently pinned down for every sum rule we have. Taking
derivative with respect to $1/M^2$ for the LCSR representations, and
adjusting the yielded sum rules for $B$-meson mass to its
measurement value, one arrives at a common result,
$s_0=(34\pm0.5)\mathrm{GeV}^{2}$, which is below the threshold value
estimated in the conventional LCSR approach, as expected. As for the
Borel parameter, we choose to use, as a sum rule window shared in
all these cases, the interval $M^2=(13-21)~\mathrm{GeV}^{2}$, in
which whereas the lower limit is obtained by keeping the twist-4
terms numerically reasonably small, the upper limit is determined by
demanding that the higher-resonance and continuum contribution
should not get too large.

\begin{figure}[]
\subfigure[]{ \label{fig:f+Mu}
\begin{minipage}[t]{0.5\textwidth}
\centering
\includegraphics[scale=0.6]{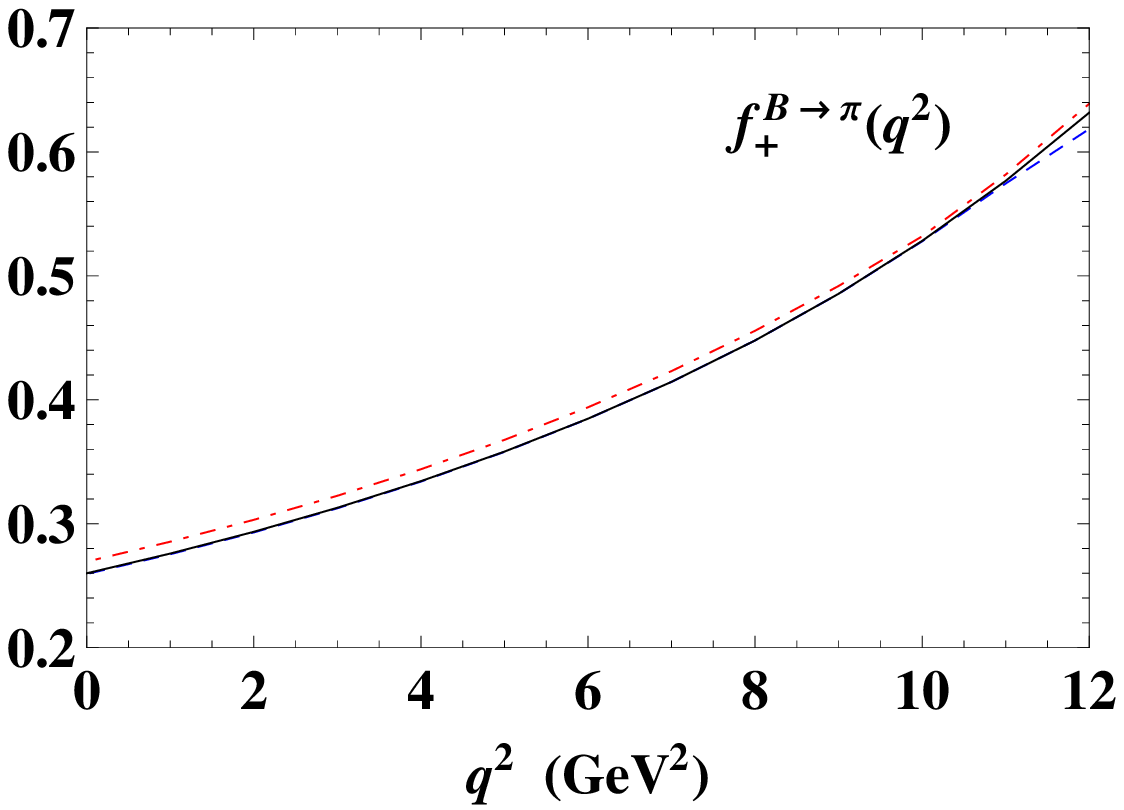}
\end{minipage}}%
\subfigure[]{ \label{fig:f0Mu}
\begin{minipage}[t]{0.5\textwidth}
\centering
\includegraphics[scale=0.6]{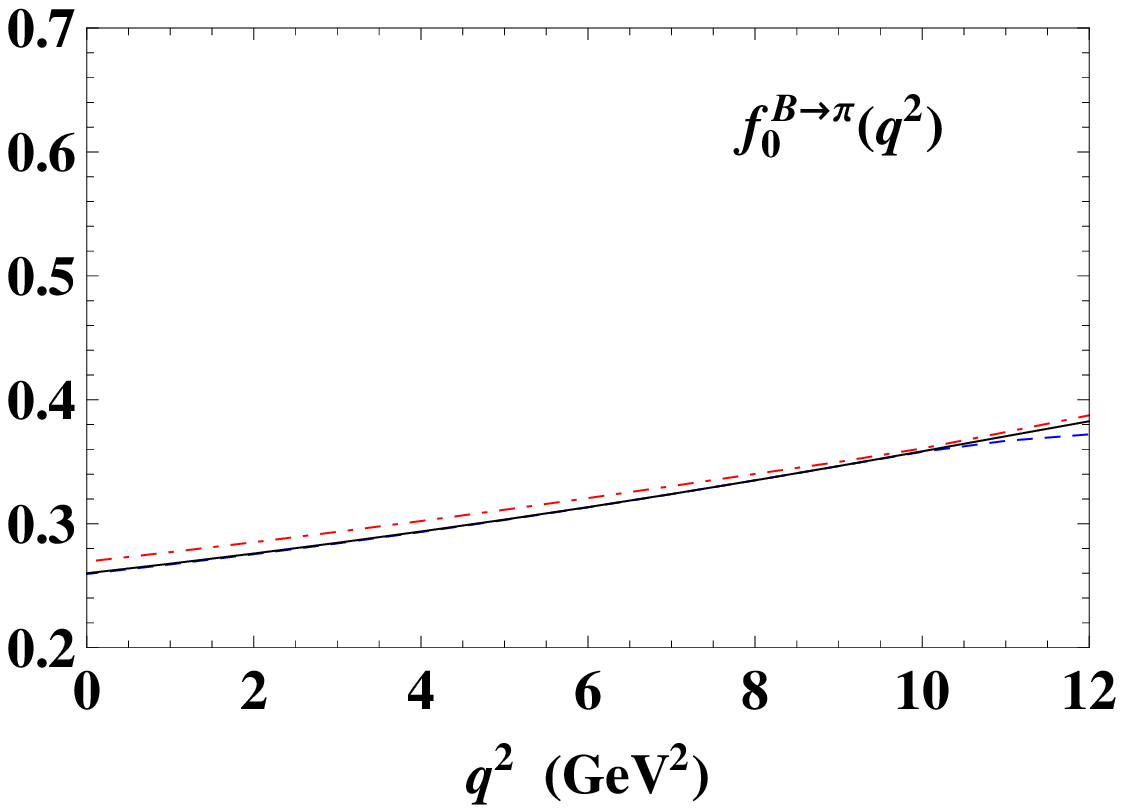}
\end{minipage}}%
\begin{center}
\caption{Stability of the LCSRs for $f_{+}^{B\to\pi}(q^2)$ and
$f_{0}^{B\to\pi}(q^2)$ with respect to the variation of the
factorization scale $\mu$. The solid lines denote the central values
and the regions between the red dot-dashed and blue dashed lines do
the uncertainties.}\label{fig:f+0Mu}
\end{center}
\end{figure}
\begin{figure}[]
\subfigure[]{ \label{fig:f+Bpi}
\begin{minipage}[t]{0.5\textwidth}
\centering
\includegraphics[scale=0.6]{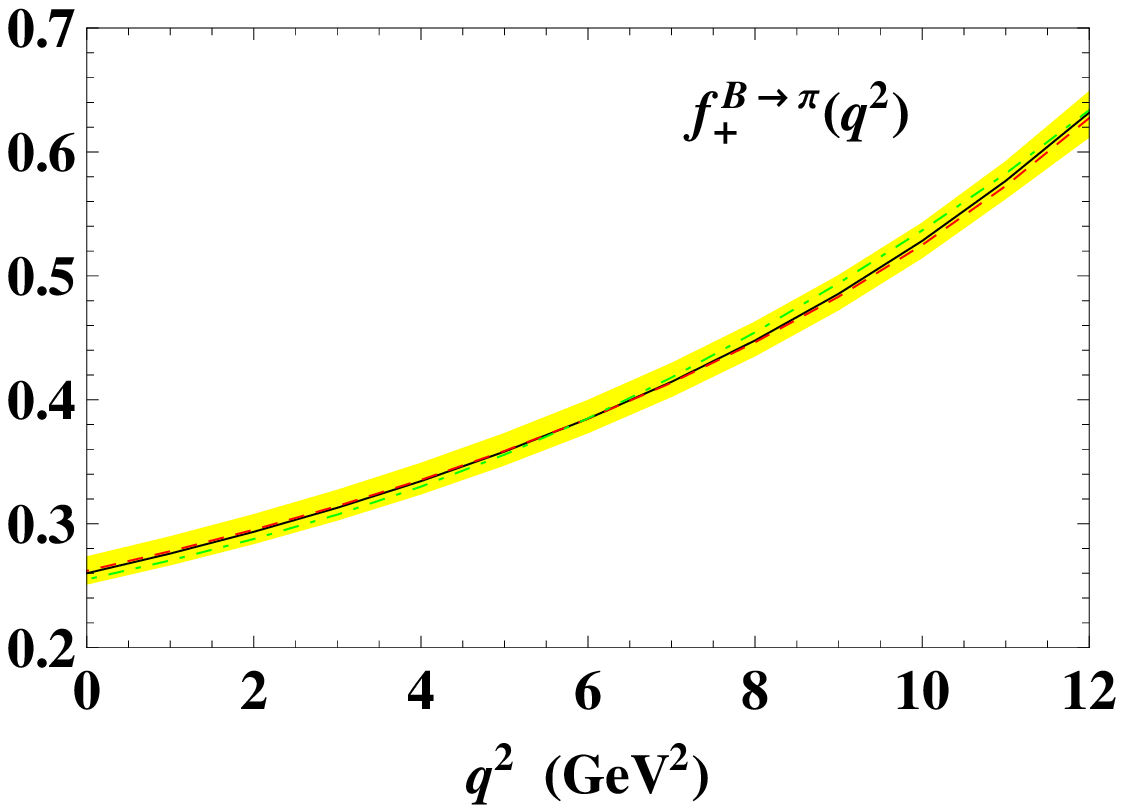}
\end{minipage}}%
\subfigure[]{ \label{fig:f0Bpi}
\begin{minipage}[t]{0.5\textwidth}
\centering
\includegraphics[scale=0.6]{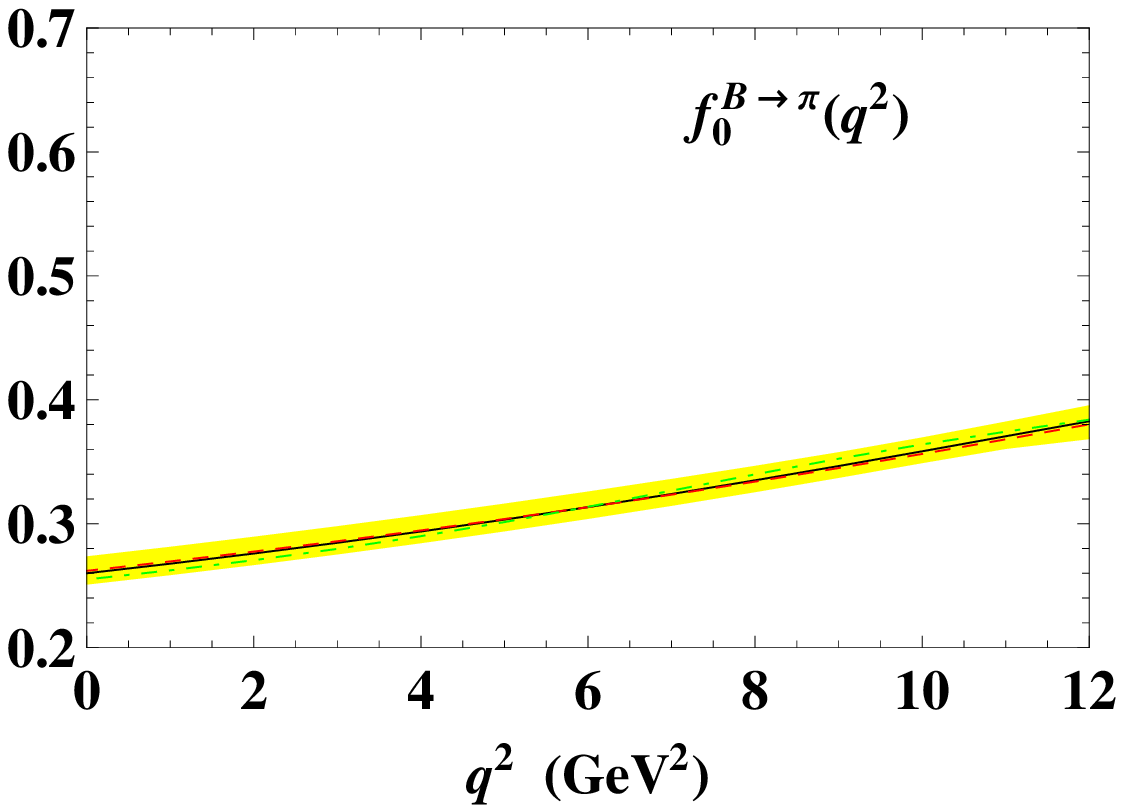}
\end{minipage}}%
\begin{center}
\subfigure[]{ \label{fig:fTBpi}
\begin{minipage}[t]{0.5\textwidth}
\centering
\includegraphics[scale=0.6]{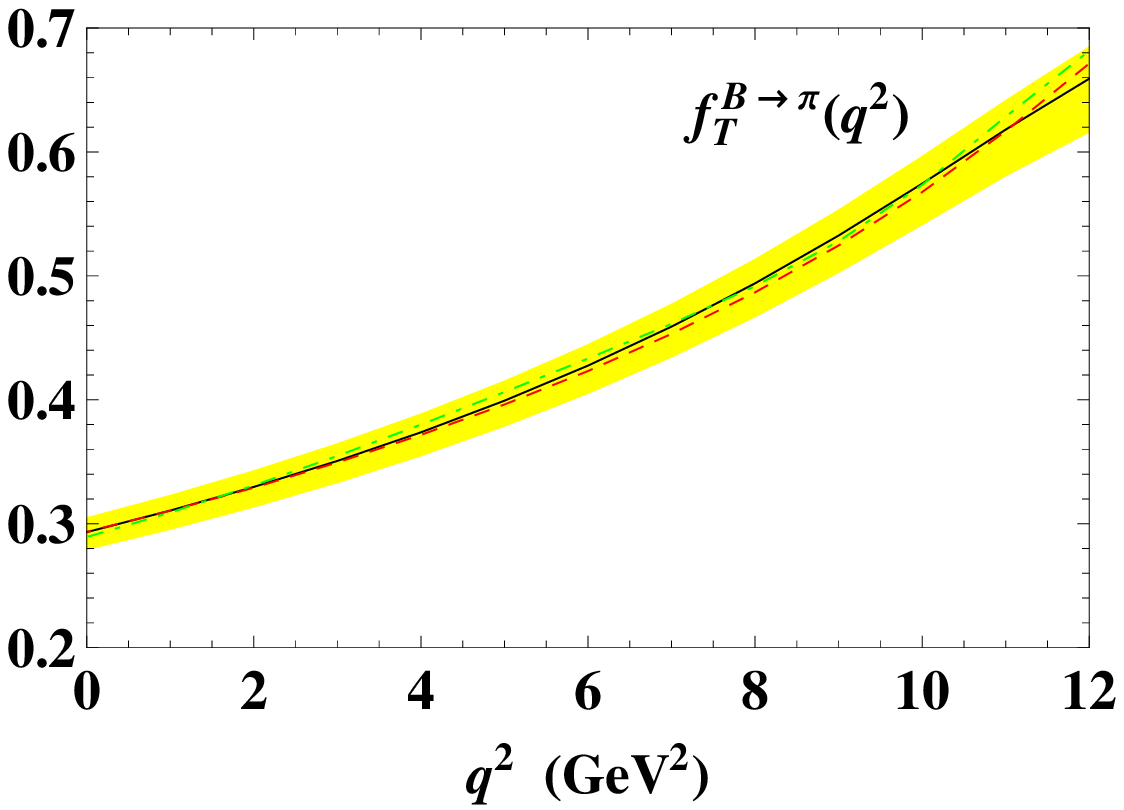}
\end{minipage}}%
\caption{$q^2$-dependence of the $B\to\pi \ell^+\ell^-$ from factors
from the LCSR. The solid lines denote the cental values and the
yellow shadow regions show the uncertainties. The green dot-dashed
(red dashed) lines denote the central values corresponding to the
parameter set $(\mathrm{\uppercase\expandafter{\romannumeral1}})$
$\left( (\mathrm{\uppercase\expandafter{\romannumeral2}})\right)$ of
\eqref{3sets}.}\label{fig:f+0TBpi}
\end{center}
\end{figure}
Equipped with the specified inputs, we can assess $q^2$-behavior of
the $B\to \pi\ell^+\ell^-$ form factors in the region $0\leq q^2\leq
12~\mathrm{GeV}^2$ we conservatively set. The consistency is
affirmed among the calculations with the three sets of fitted moment
parameters. We intend to illustrate our numerical results by
focusing on the case with the parameter set
$(\mathrm{\uppercase\expandafter{\romannumeral3}})$ in
\eqref{3sets}. There is a good stability of the sum rules against
the variation of the Borel parameter (see Fig.\ref{fig:f+0TMM}).
Moreover, as the scale parameter varies in the interval required,
the sum rule results for $f_{+}^{B\to\pi}(q^2)$ and
$f_{0}^{B\to\pi}(q^2)$ change by respectively less than $4.0\%$ and
$3.9\%$ depending on $q^2$, showing less sensitivity to that
parameter, as expected and shown in Fig.\ref{fig:f+0Mu}. Including
the uncertainties achieved by adding in quadrature the separate
errors due to variations of the inputs, our predictions for the form
factors, which are compatible with the corresponding those using the
standard LCSR approach \cite{LSCR-A.Khodja-0801.1796,
LSCR-A.Khodja-1103.2655}, are displayed in Fig.\ref{fig:f+0TBpi}.
The results at $q^{2}=0$ read,
\begin{eqnarray}
&&f_{+}^{B\to\pi}(0)=f_{0}^{B\to\pi}(0)=0.260^{+0.013}_{-0.008},\\
&&f_{T}^{B\to\pi}(0)=0.293^{+0.011}_{-0.014}.
\end{eqnarray}

It is not hard to analytically continue the sum rule predictions to
the high-$q^2$ region in a certain form factor parametrization. This
has been done in \cite{Y.Wang-1407.2468}, where the form factors
obtained at $\mu=4.8~\mathrm{GeV}$ and extrapolated analytically
were used to confine the Majorana neutrino contribution to the
$B\to\pi \mu^+\mu^- $, aimed at studying the same-sign dilepton
decays of $B$ meson induced by such neutrino, a lepton flavor
violating channel. From Fig.\ref{fig:fTBpi} we can see, though, that
when the distinct sets of moment parameter inputs are taken, the
resulting shapes of the tensor form factor, despite being very close
to one another, have the different trends of evolution to high
$q^2$. This would make uncertainty in the extrapolation large. On
the other hand and more importantly, for having a reliable
parametrization it is necessary to use some available estimates of
these form factors at large $q^2$'s as an additional bound on their
behavior in the entire kinematically allowed time-like region.
\section{Form factor shapes in the whole kinematical region}
Estimating the form factors in the $q^2$ region accessible for
$B\to\pi \ell^+\ell^-$, $0\lesssim q^{2}\leq (m_{B}-m_{\pi})^{2}$
(the upper limit is about $26.4~\mathrm{GeV^2}$ with the measured
pion mass listed in Tab.~\ref{tab:InPut}), we make the best of their
analyticity, as well as the results based on the LCSR approach and
LQCD simulation (or less model-dependent assumption). To be
specific, for each form factor we fit simultaneously the theoretical
predictions of these approaches to a series expansion in the mapping
function $z(q^{2},t_{0})$, which transforms the complex $q^2$-plane
with a cut along the positive real axis onto the inner part of the
unit circle $\mid z \mid=1$ in the $z$-plane
\cite{BGL-9412324,BCL-0807.2722},
\begin{eqnarray}
z(q^{2},t_{0})=\frac{\sqrt{(m_{B}+m_{\pi})^{2}-q^{2}}-\sqrt{(m_{B}+m_{\pi})^{2}-t_{0}}}{\sqrt{(m_{B}+m_{\pi})^{2}-q^{2}}+\sqrt{(m_{B}+m_{\pi})^{2}-t_{0}}},
\end{eqnarray}
where the auxiliary parameter $t_{0}(<(m_{B}+m_{\pi})^{2})$ can be
chosen as
$t_{0}=(m_{B}+m_{\pi})^{2}-2\sqrt{m_{B}m_{\pi}}\sqrt{(m_{B}+m_{\pi})^{2}-q_{0}^{2}}$.
Because the kinematical region in consideration can be mapped onto a
quite small interval in the $z$ plane by selecting optimally
$t_{0}$, an expansion around $z=0$, with the first few terms
retained, furnishes a state-of-the-art analytic approach to the form
factors.

Taking into account general analytic properties of the form factors,
instead of the direct expansion we adopt the
Bourrely-Caprini-Lellouch (BCL) versions
\cite{BCL-0807.2722,LSCR-DK-Khodjamirian-0907.2842}:
\begin{eqnarray}\label{eq:BCL-f+0}
f_{+(T)}^{B\to\pi}(q^{2})&=&\frac{f_{+(T)}^{B\to\pi}(0)}{1-q^{2}/m_{B^{*}}^{2}}
\left\{1+\sum_{k=1}^{N-1}b_{k}^{+(T)}
\bigg[z(q^{2},t_{0})^{k}-z(0,t_{0})^{k}\right.\nonumber\\
&&\left.\left.-(-1)^{N-k}\frac{k}{N}\left(z(q^{2},t_{0})^{N}-z(0,t_{0})^{N}\right)\right]\right\},\\\label{eq:BCL-fT}
f_{0}^{B\to\pi}(q^{2})&=&f_{0}^{B\to\pi}(0)\left\{1+\sum_{k=1}^{N}b_{k}^{0}
\left(z(q^{2},t_{0})^{k}-z(0,t_{0})^{k}\right)\right\},
\end{eqnarray}
with the $B^*$-meson mass $m_{B^*}=5.325~\mathrm{GeV}$. Having a
prefactor $\sim(1-q^{2}/m_{B^{*}}^{2})^{-1}$, the $z$ series
\eqref{eq:BCL-f+0} provides an improvement to the $B^*$-pole
dominance. For the scalar form factor there is no similar factor
involved for obvious reason. The condition of unitarity does not
provide, for small $N$, a restrictive bound on the coefficients
$b_k^{+(0,T)}$, as argued in \cite{BCL-0807.2722}. For simplicity we
choose to work with a two-parameter form; that is, we truncate the
expansions \eqref{eq:BCL-f+0} and \eqref{eq:BCL-fT} by taking $N=3$
and $N=2$, respectively.

At present, the LCSR predictions for both $f_{+}^{B\to\pi}(q^2)$ and
$f_{0}^{B\to\pi}(q^2)$ presented in the above section, along with
the LQCD results from the HPQCD \cite{LQCD-f+0HPQCD-0601021} and
FNAL/MILC \cite{LQCD-Bpif+0MILC-0811.3640} collaborations are
applicable to constrain the expansion parameters $b_i^{+(0)}$
(i=1,2). However, taking into account the fact that the $b$ quark in
the correlation function is even farther away from its mass shell at
$q^{2}<0$ than at $q^{2}>0$, instead of the form factor shapes
obtained in the interval $0\leq q^{2}\leq 12~\mathrm{GeV^2}$ we
employ the LCSR results in an enlarged region $q_0^2\leq q^{2}\leq
12~\mathrm{GeV^2}(q_0^2<0)$ in the numerical fitting, to put more
restraints on the expansion coefficients. The lower limit $q_0^2$ is
specified as $q_0^2=-5~\mathrm{GeV^2}$, in order to make the
light-cone OPE's have a good perturbative hierarchy. With this
choice, we could fix the parameter $t_0$ and further the mapping
function. A couple of typical mapping results are listed as follows:
$q_0^2\to |z|=0.30$, $q^2=0\to |z|=0.26$ and
$q^2=26.4~\mathrm{GeV}^2\to |z|=0.13$. In addition, to enhance
fitting precision use should be made of the sum rule for the ratio
$f_{+(0)}^{B\to\pi}(q^{2})/f_{+(0)}^{B\to\pi}(0)$, in which the
uncertainties due to the nonperturbative inputs cancel out in part.

The fitting is done in the standard procedure. The optimal parameter
sets, yielded for each of the vector and scalar form factors by
separately fitting the central values, upper and lower limits of the
LCSR and LQCD results, are summarized in Tab.~\ref{tab:f+0b1b2}.
Using these as input, we get an analytical expression for describing
$q^2$ dependence of both form factors in the kinematically allowed
space-like as well as time-like regions.
\begin{table}[h]
\centering \caption {BCL parameter sets obtained for separate
estimates for the central values, upper and lower limits of the
$B\to \pi$ vector (scalar) form factor as a function of
$q^2$.}\label{tab:f+0b1b2}
\def\temptablewidth{0.9\textwidth}
\begin{tabular*}{\temptablewidth}{@{\extracolsep{\fill}}c|ccc}
\toprule[1.0pt]
\toprule[1.0pt]
 Parameter sets &Central values  &Upper limits  &Lower limits\\
\hline
 $(f_{+}^{B\to\pi}(0),b_{1}^{+},b_{2}^{+})$ &(0.260, $-$2.357, $-$1.411) &(0.273, $-$3.124, 1.195) &(0.252, $-$1.451, $-$4.099)\\
 $(f_{0}^{B\to\pi}(0),b_{1}^{0},b_{2}^{0})$ &(0.260, $-$6.933, 6.635)  &(0.273, $-$7.389, 8.351) &(0.252, $-$5.828, 4.600)\\
\bottomrule[1.0pt]
\bottomrule[1.0pt]
\end{tabular*}
\end{table}

For the tensor form factor, a lattice simulation has recently been
performed \cite{C.M.Bouchard-1310.3207}, but only a preliminary
result being known. To arrive at a good understanding of its
behavior at high $q^2$, the authors of \cite{BPill-A.Ali-1312.2523}
connected the $B\to\pi$ with the corresponding $B\to K$ form
factors,
\begin{eqnarray}
f^{B\to \pi}_{+(0,T)}(q^2)=\frac{f^{B\to
K}_{+(0,T)}(q^2)}{1+R_{+(0,T)}(q^2)},
\end{eqnarray}
by invoking a $SU_F(3)$ symmetry breaking function
$R_{+(0,T)}(q^2)$, and made use of the available lattice results on
both $f^{B\to K}_{+,0,T}(q^2)$ \cite{LQCD-BKfT-1306.2384} and
$f^{B\to\pi}_{+,0}(q^2)$ and an ansatz
\begin{eqnarray}
R_{T}(q^2)=\frac{R_{+}(q^2)+R_{0}(q^2)}{2},
\end{eqnarray}
which has been shown to be effective for low $q^2$ by a calculation
based on heavy quark symmetry \cite{BPill-A.Ali-1312.2523}. The
results at eight $q^2$'s are demonstrated in Fig.5(c). We need to
stress that they are obtained at $\mu=4.8~\mathrm{GeV}$, the same
scale at which the low-$q^2$ behavior of the form factor has been
predicted before by resorting to the LCSR method. Seeing that the
resulting predictions may be regarded as QCD-based to a large
extent, it should be in order that we use them for the parameter
fitting in the absence of an available lattice estimate. Then the
BCL parameter sets for the tensor form factor are accessible by
doing the same as in the case of $f^{B\to \pi}_{+(0)}(q^2)$. We give
the best-fitted results in Tab.~\ref{tab:fTb1b2}.

\begin{table}[h]
\centering \caption {BCL parameter sets obtained for separate
estimates for the central values, upper and lower limits of the
$B\to \pi$ tensor form factor as a function of
$q^2$.}\label{tab:fTb1b2}
\def\temptablewidth{0.9\textwidth}
\begin{tabular*}{\temptablewidth}{@{\extracolsep{\fill}}c|ccc}
\toprule[1.0pt]
\toprule[1.0pt]
 Parameter sets &Central values  &Upper limits  &Lower limits\\
\hline
 $(f_{T}^{B\to\pi}(0),b_{1}^{T},b_{2}^{T})$ &(0.293, $-$0.821, $-$2.266) &(0.304, $-$1.136, $-$1.957) &(0.279, $-$0.534, $-$2.509)\\
\bottomrule[1.0pt]
\bottomrule[1.0pt]
\end{tabular*}
\end{table}
\begin{figure}[]
\subfigure[]{ \label{fig:f+}
\begin{minipage}[t]{0.5\textwidth}
\centering
\includegraphics[scale=0.6]{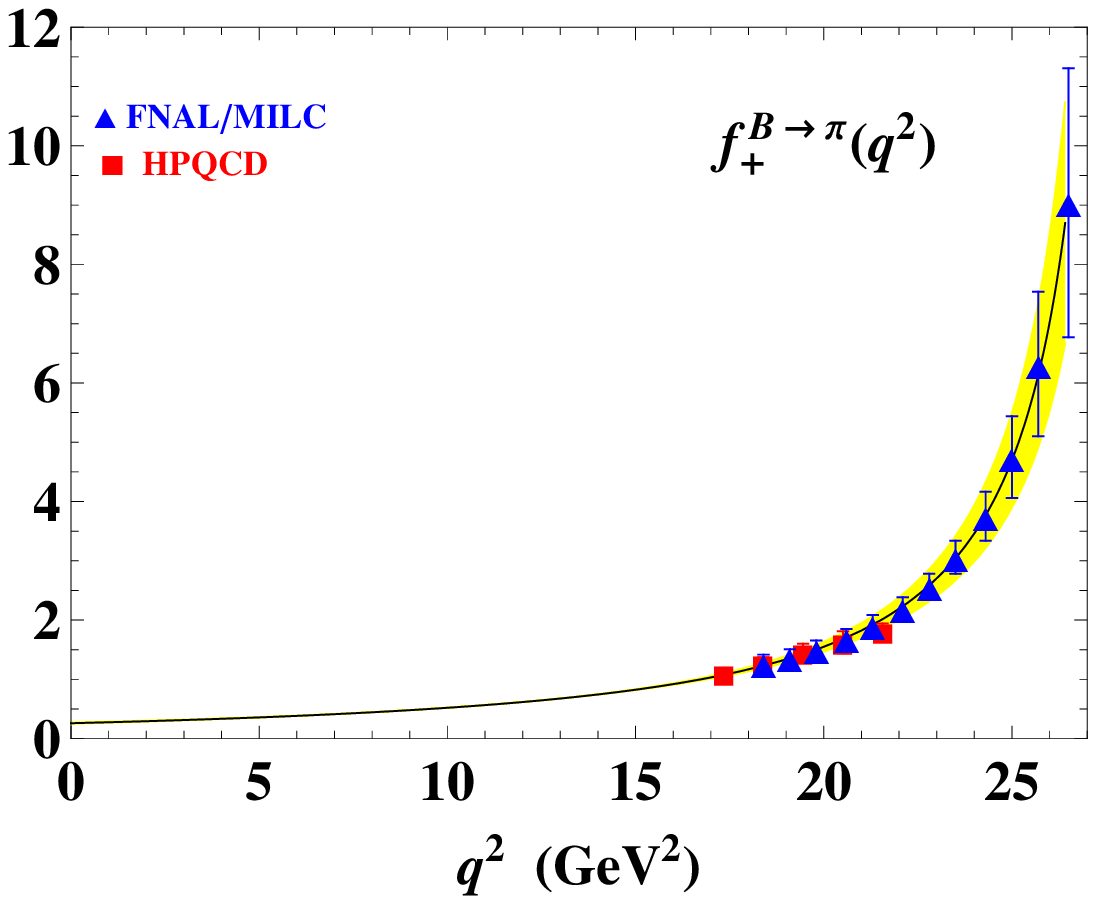}
\end{minipage}}%
\subfigure[]{ \label{fig:f0}
\begin{minipage}[t]{0.5\textwidth}
\centering
\includegraphics[scale=0.6]{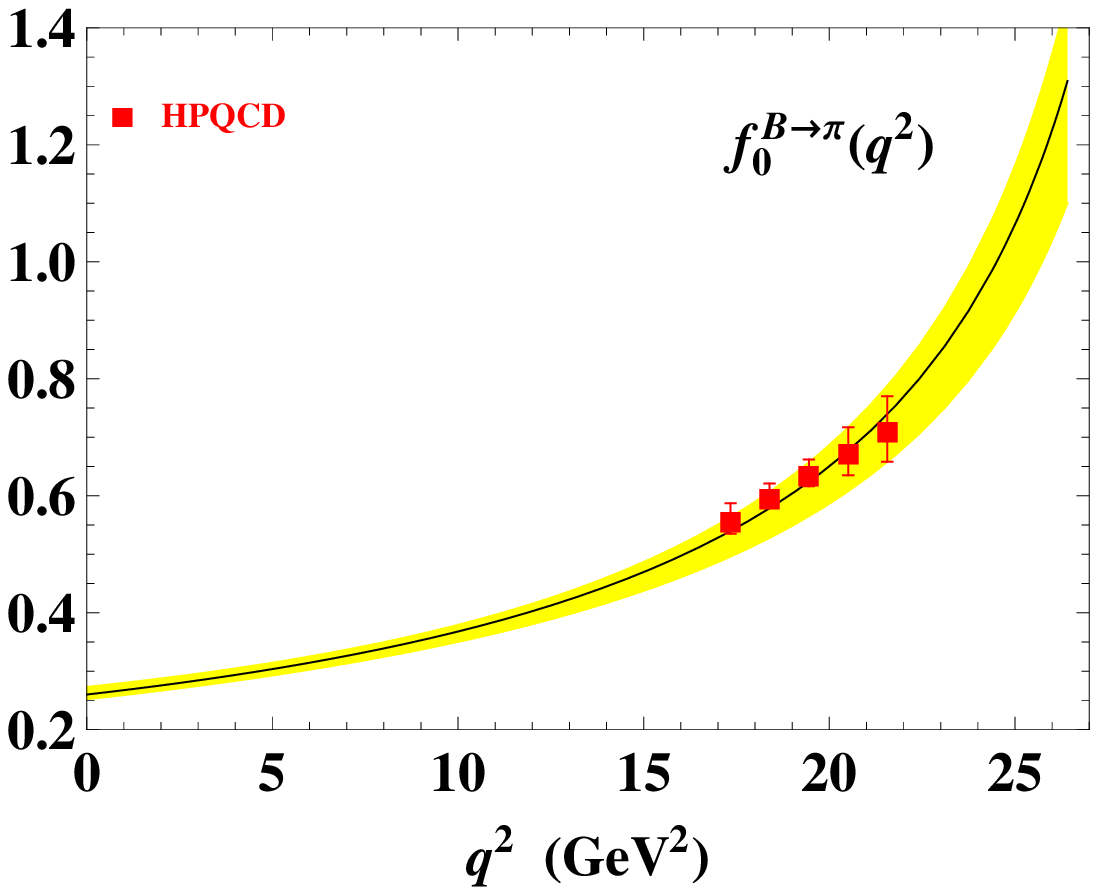}
\end{minipage}}%
\begin{center}
\subfigure[]{ \label{fig:fT}
\begin{minipage}[t]{0.5\textwidth}
\centering
\includegraphics[scale=0.6]{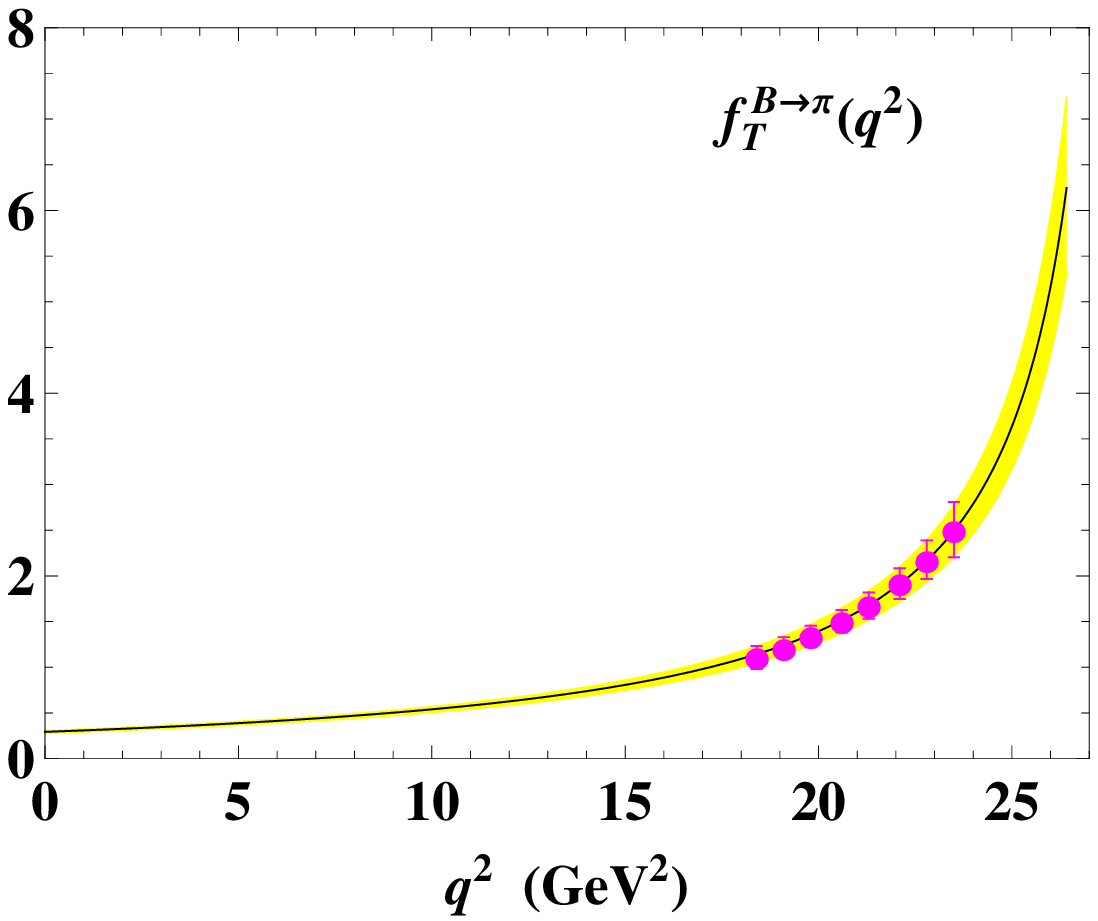}
\end{minipage}}%
\vspace{-0.3cm} \caption{Shapes of the $B\to\pi \ell^+\ell^-$ from
factors from the fitted BCL parameterizations. The solid lines
represent the cental values and the yellow shadow regions do the
uncertainties. The magenta vertical-bars in (c) indicate the results
based on both the LQCD data on the $B\to K$ tensor form factor and
the $SU_F(3)$ symmetry breaking ansatz.}\label{fig:f+0T}
\end{center}
\end{figure}
In the fitted BCL parameterizations \eqref{eq:BCL-f+0} and
\eqref{eq:BCL-fT}, the $B\to \pi \ell^+\ell^-$ form factors are now
understandable in the whole $q^2$ region. Illustrated are the
resulting shapes of the vector, scalar and tensor form factors,
respectively, in Figs. \ref{fig:f+0T} (a), (b) and (c). At the zero
recoiling point $q^2=26.4~\mathrm{GeV}^2$, there are the following
observations: $f_{+}^{B\to\pi}(q^2=26.4~\mathrm{GeV}^2)=8.630$,
$f_{0}^{B\to\pi}(q^2=26.4~\mathrm{GeV}^2)=1.306$ and
$f_{T}^{B\to\pi}(q^2=26.4~\mathrm{GeV}^2)=6.203$. At this point, let
us make a simple comparison between the present findings and those
of \cite{BPill-A.Ali-1312.2523}. The behavior we predict for
$f_{+}^{B\to\pi}(q^2)$ resembles closely the one given in that
literature through fitting the measured shape of the form factor
multiplied by the CKM matrix element $|V_{ub}|$, which provides a
theoretical interpretation for the observation based on
data-fitting. For both $f_{0}^{B\to\pi}(q^2)$ and
$f_{T}^{B\to\pi}(q^2)$, a consistent result is also observed within
the estimated errors. However, whereas in
\cite{BPill-A.Ali-1312.2523} heavy quark symmetry is applied to
constrain behavior of these two form factors in the large recoil
region, here we employ for the same purpose the LCSR calculations
with a chiral current correlator, from which the resulting
heavy-to-light form factors could comply explicitly with the heavy
quark limit behavior as predicted by soft collinear effective theory
(SCET) \cite{C.W.Bauer-0011336}, having the symmetry breaking
corrections included systematically in the present estimates.

Certainly our theoretical predictions are available to
systematically study the semileptonic $B\to \pi$ decays, including
the tau lepton modes which are sensitive to the extensions of the SM
with several Higgs fields. Moreover, the same approach as above
applies to exploring the $B \to K\ell^+\ell^-$ rare processes. These
discussions, nevertheless, are beyond the scope of this paper.

\section{Dilepton invariant mass distributions in $B\to \pi\ell^+\ell^-$ and branching ratios}

Having in hand the $B\to \pi$ form factor parameterizations based on
the available LQCD (or the quasi-model independent) and LCSR
estimates obtained respectively at small and large recoil regions,
we can make predictions on the decay rates and branching fractions
for $B\to \pi \ell^+\ell^-$ within the naive factorization
framework. The matrix elements for $B\to \pi \ell^+\ell^-$ are
written as \cite{BXsll-A.J.Buras-9501281,BXdll-F.Kruger-9608361}
\begin{eqnarray}\label{eq:Amplitude}
\mathcal{M}&=&\frac{G_{F}\alpha_{\mathrm{em}}}{\sqrt{2}\pi}V_{tb}V^{*}_{td}\bigg[C_{9}^{\mathrm{eff}}(\mu)\langle\pi(p)|\bar{d}\gamma_{\mu}P_Lb|B(p+q)\rangle\bar{\ell}\gamma^{\mu}\ell\nonumber\\
&&+C_{10}^{\mathrm{eff}}(\mu)\langle \pi(p)|\bar{d}\gamma_{\mu}P_Lb|B(p+q)\rangle\bar{\ell}\gamma^{\mu}\gamma_{5}\ell)\nonumber\\
&&\left.-2C_{7}^{\mathrm{eff}}(\mu)\frac{m_b}{q^{2}}\langle
\pi(p)|\bar{d}i\sigma_{\mu\nu}q^{\nu}P_R
b|B(p+q)\rangle\bar{\ell}\gamma^{\mu}\ell\right],
\end{eqnarray}
based on the effective Hamiltonian responsible for the $b\to
d\ell^{+}\ell^{-}$ transitions
\cite{BXsll-A.J.Buras-9501281,G.Buchalla-9512380}:
\begin{eqnarray}\label{eq:Hamiltonian}
\mathcal{H}_{eff}&=&-\frac{4G_{F}}{\sqrt{2}}V_{tb}V_{td}^{*}\left[\sum_{i=1}^{10}C_{i}(\mu)O_{i}(\mu)-\lambda_{u}\sum_{i=1}^{2}C_{i}(\mu)\left(O_{i}^{u}(\mu)-O_{i}(\mu)\right)\right].
\end{eqnarray}
Here $G_{F}$ and $\alpha_{\mathrm{em}}$ are respectively the Fermi
coupling and the fine structure constants, $P_{L,R}=(1\mp
\gamma_5)/2$, $\lambda_{u}(=V_{ub}V_{ud}^{*}/V_{tb}V_{td}^{*})$ is
of the standard parameteriztion form,
\begin{eqnarray}
\lambda_{u}=\frac{\bar{\rho}-i\bar{\eta}}{1-\bar{\rho}+i\bar{\eta}},
\end{eqnarray}
subjected to a minor correction of $\mathcal{O}(\lambda^{5})$
$(\lambda=|V_{us}|)$, $O_{i}(\mu)$ denote the dimension-six
operators with the corresponding Wilson coefficients $C_{i}(\mu)$,
and $C^{\mathrm{eff}}_{7,9,10}(\mu)$ stand for the effective Wilson
coefficients which are particular combinations of $C_i(\mu)$ and
given by \cite{BXsr-A.J.Buras-9311345,BKll-A.J.Buras-0811.1214},
\begin{eqnarray}
&&C_7^{\mathrm{eff}}=\frac{4\pi}{\alpha_s}C_7-\frac{1}{3}C_3-\frac{4}{9}C_4-\frac{20}{3}C_5-\frac{80}{9}C_6,\\\label{eq:C9eff}
&&C_9^{\mathrm{eff}}(q^2)=\frac{4\pi}{\alpha_s}C_9+\frac{4}{3}C_3+\frac{64}{9}C_5+\frac{64}{27}C_6+Y_{\mathrm{SD}}(q^2)+Y_{\mathrm{LD}}(q^2),\\
&&C_{10}^{\mathrm{eff}}=\frac{4\pi}{\alpha_s}C_{10}.
\end{eqnarray}
As shown in \eqref{eq:C9eff}, $C_9^{\mathrm{eff}}(q^2)$ depends on $q^2$ through
the dynamical functions $Y_{\mathrm{SD}}(q^2)$ and
$Y_{\mathrm{LD}}(q^2)$ parameterizing, respectively, the short-and
the long-distance contributions due to the four-quark operators. The
former is obtained as
\begin{eqnarray}
&&Y_{\mathrm{SD}}(q^2)=h(q^2,m_c)\left(\frac{4}{3}C_1+C_2+6C_3+60C_5\right)\nonumber\\
&&~~~~~~~~~~-h(q^2,m_b)\left(\frac{7}{2}C_3+\frac{2}{3}C_4+38C_5+\frac{32}{3}C_6\right)\nonumber\\
&&~~~~~~~~~~-h(q^2,0)\left(\frac{1}{2}C_3+\frac{2}{3}C_4+8C_5+\frac{32}{3}C_6\right)\nonumber\\
&&~~~~~~~~~~+\lambda_u\left(\frac{4}{3}C_1+C_2\right)\left(h(q^2,m_c)-h(q^2,0)\right),
\end{eqnarray}
where $h(q^2,m_q)$ is the loop function dependent on the mass
parameter $m_q$, which indicates the quark pole mass as $q=c,b$ and
has been set to zero for the light quarks,
\begin{eqnarray}
h(q^2,m_q)=-\frac{4}{9}\left(\mathrm{ln}\frac{m_q^2}{\mu^2}-\frac{2}{3}-x\right)-\frac{4}{9}(2+x)\sqrt{|x-1|}\times
\left\{ \begin{aligned}
            &\mathrm{ln}\frac{1+\sqrt{1-x}}{\sqrt{x}}-i\frac{\pi}{2},~x\leq1\\
            &\mathrm{arctan}\frac{1}{\sqrt{x-1}},~~~~~~~~x>1
        \end{aligned} \right.
\end{eqnarray}
with $x=4m_q^2/q^2$. The Wilson coefficients have been computed in
next-to-next-to-leading logarithmic (NNLL) approximation through a
two-loop matching of the effective with the full theory at the scale
of the $W$-boson mass, and then evolved down to $\mu\sim m_b$ with
the aid of the QCD renormalization group equations. The SM values at
$\mu=4.8\mathrm{GeV}$ are given in Tab.~\ref{tab:WilsonC}. To
determine $q^2$-dependence of $Y_{\mathrm{SD}}(q^2)$, we make use of
the same inputs as in \cite{BPill-A.Ali-1312.2523} for the quark
pole masses, $m_b=4.91~\mathrm{GeV}$ and $m_c=1.77~\mathrm{GeV}$,
following from a three-loop QCD calculation
\cite{3loopQCD-K.Chetyrkin-9907509,3loopQCD-K.Chetyrkin-9911434,
3loopQCD-K.Chetyrkin-9912391} with an additional electro-weak
correction \cite{weakCorrection-Z.Z.Xing-0712.1419} included. The
long-distance resonance dynamics embedded in $Y_{\mathrm{LD}}(q^2)$
is not in our consideration, because in the resonant regions such
contributions can be removed experimentally and it may generally be
believed, on the basis of the evaluation made for the corresponding
$B\to K $ decays \cite{LCSR-K-A.Khodjamirian-1211.0234} and some
indirect experimental observations, that beyond these regions the
hadron resonances bring about only a moderate impact on the width.
\begin{table}[h]
\centering \caption {SM (effective) Wilson coefficients at
$\mu=4.8~\mathrm{GeV}$, in next-to-next-to-leading logarithmic
(NNLL) approximation \cite{BKll-A.J.Buras-0811.1214}. The function
$Y(q^2)$ is defined as
$Y(q^2)=\frac{4}{3}C_3+\frac{64}{9}C_5+\frac{64}{27}C_6+Y_{\mathrm{SD}}(q^2)+Y_{\mathrm{LD}}(q^2)$.}\label{tab:WilsonC}
\def\temptablewidth{0.9\textwidth}
\begin{tabular*}{\temptablewidth}{@{\extracolsep{\fill}}cccccccccc}
\toprule[1.0pt]
\toprule[1.0pt]
 $C_{1}$ &$C_{2}$ &$C_{3}$ &$C_{4}$ &$C_{5}$
 &$C_{6}$ &$C_{7}^{\mathrm{eff}}$ &$C_{9}^{\mathrm{eff}}-Y(q^2)$ &$C_{10}^{\mathrm{eff}}$\\
 $-0.257$ &$1.009$ &$-0.005$ &$-0.078$ &$0.000$
 &$0.001$ &$-0.304$ &$4.211$ &$-4.103$ \\
\bottomrule[1.0pt]
\bottomrule[1.0pt]
\end{tabular*}
\end{table}

For the $b\to d $ hadronic matrix elements entering
\eqref{eq:Amplitude} using their standard parameterization forms
\eqref{eq:MatrixElements}, we get the differential branching ratios
for, say, the charged decay models $B^-\to \pi^-\ell^{+}\ell^{-}$,
\begin{eqnarray}\label{eq:Branching}
\frac{d\mathcal{B}(B^-\to \pi^-\ell^{+}\ell^{-})}{dq^2}
=\frac{G_{F}^{2}\alpha^{2}_{\mathrm{em}}\tau_{B}}{2^{10}\pi^{5}m_{B}^{3}}|V_{tb}V^{*}_{td}|^{2}\sqrt{\lambda(q^2)
\left(1-\frac{4m^2_{\ell}}{q^2}\right)}\sigma(q^2),
\end{eqnarray}
where $\tau_{B}=1.638\pm0.004~\mathrm{ps}$ denotes the $B$ meson lifetime, $m_{\ell}$ the
lepton mass, $\lambda(q^2)=(m_B^2+m_{\pi}^2-q^2)^2-4m_B^2m_{\pi}^2$,
and
\begin{eqnarray}
&&\sigma(q^2)=\frac{2}{3}\lambda(q^2)\left[\left(1+\frac{2m^2_{\ell}}{q^2}\right)\left|
C_9^{\mathrm{eff}}(q^2)f_{+}^{B\to\pi}(q^2)+\frac{2m_b}{m_B+m_{\pi}}C_7^{\mathrm{eff}}f_{T}^{B\to\pi}(q^2)\right|^2\right.\nonumber\\
&&~~~~~~~\left.+\left(1-\frac{4m^2_{\ell}}{q^2}\right)\left|
C_{10}^{\mathrm{eff}}f_{+}^{B\to\pi}(q^2)\right|^2\right]+\frac{4m^2_{\ell}}{q^2}(m_B^2-m_{\pi}^2)^2\left|
C_{10}^{\mathrm{eff}}f_{0}^{B\to\pi}(q^2)\right|^2,
\end{eqnarray}
of which the last term on the right-hand side includes a factor of
$m_{\ell}^2$ which is numerically close to zero for $\ell=e,\mu$, so
that the scalar form factor plays a negligible role for the decays
with a dielectron or dimuon in the final state.
\begin{table}[ht]
\begin{center}
\caption{Some of the parameter inputs used in the numerical analysis
\cite{PDG2014}.}\label{tab:InPut}
\def\temptablewidth{0.9\textwidth}
\begin{tabular*}{\temptablewidth}{@{\extracolsep{\fill}}cccc}
\toprule[1.0pt]
\toprule[1.0pt]
    $\alpha_{em}$   & $1/137$
    & $\bar{\eta}$  & $0.354\pm0.015$\\
    $G_{F}$         & $1.16638\times10^{-5}~\mathrm{GeV}^{-2}$
    & $m_{e}$       & $0.511~\mathrm{MeV}$\\
    $|V_{tb}|$      & $0.99914$
    & $m_{\mu}$     & $0.106~\mathrm{GeV}$\\
    $|V_{td}|$      & $0.00886^{+0.00033}_{-0.00032}$
    & $m_{\tau}$    & $1.777~\mathrm{GeV}$\\
    $\bar{\rho}$       & $0.124\pm0.024$
    & $m_{\pi}$     & $0.13957~\mathrm{GeV}$\\
\bottomrule[1.0pt]
\bottomrule[1.0pt]
\end{tabular*}
\end{center}
\end{table}
\begin{figure}[!ht]
\begin{center}
\includegraphics[scale=0.8]{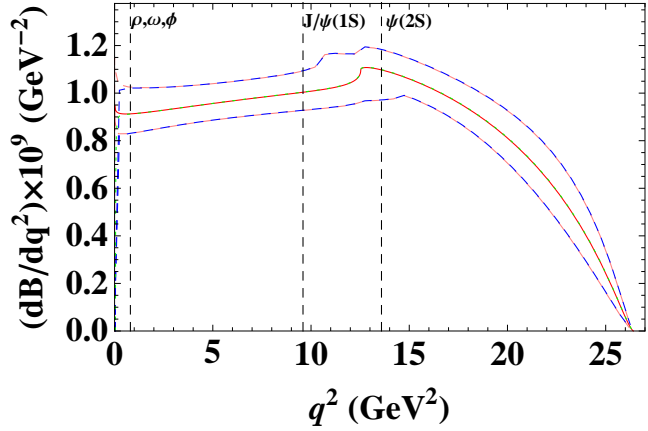}
\caption{Invariant mass distributions in $B^{-}\to\pi^{-}e^{+}e^{-}$
and $B^{-}\to\pi^{-}\mu^{+}\mu^{-}$, with the central values denoted
by respectively the red solid and green dotted lines, and the
uncertainty bounds shown by respectively the pink dashed and blue
dashed lines.}\label{fig:dBSMeMu}
\end{center}
\end{figure}
\begin{table}[h]
\centering \caption {Summary of the SM predictions (in $10^{-8}$)
for the branching ratios for $B^-\to\pi^-\ell^{+}\ell^{-}$
($\ell=e,\mu,\tau$) in naive factorization.}\label{tab:Br}
\begin{tabular}{c|c|ccccc}
\toprule[1.0pt] \toprule[1.0pt]
 \multirow{2}{*}{Modes}&\multirow{2}{*}{LHCb experiment\cite{BPi-Observation}}
&\multicolumn{5}{c}{Theoretical predictions}\\
\cline{3-7} &&This work  &\cite{BPill-J.J.Wang-0711.0321}
&\cite{BPill-Z.J.Xiao-1207.0265}
 &\cite{BPill-A.Ali-1312.2523} &\cite{Bpill-R.N.Faustov-1403.4466}\\
\midrule[0.7pt]
 $B^-\to\pi^- e^{+}e^{-}$ &--\,-- &$2.263^{+0.227}_{-0.192}$  &$2.03\pm0.23$ &$1.95^{+0.61}_{-0.48}$
 &--\,-- &--\,--\\
 $B^-\to\pi^-\mu^{+}\mu^{-}$  &$2.3\pm0.6\pm0.1$ &$2.259^{+0.226}_{-0.191}$  &$2.03\pm0.23$ &$1.95^{+0.61}_{-0.48}$
 &$1.88^{+0.32}_{-0.21}$ &$2.0\pm0.2$\\
 $B^-\to\pi^-\tau^{+}\tau^{-}$  &--\,-- &$1.017^{+0.118}_{-0.139}$  &--\,-- &$0.60^{+0.18}_{-0.14}$
 &--\,-- &$0.70\pm0.07$\\
\bottomrule[1.0pt] \bottomrule[1.0pt]
\end{tabular}
\end{table}
\begin{table}[h]
\centering \caption{Theoretical predictions for the
$B^-\to\pi^-\mu^{+}\mu^{-}$ partial branching ratios in some chosen
$q^2$ intervals (in $10^{-8}$).}\label{tab:PartialBr}
\label{tab:BpillInPut}
\begin{tabular}{|c|ccc|}
\hline
 \multirow{2}{*}{~~~~~$[q_{min}^{2},q_{max}^{2}]$~~~~~} & \multicolumn{3}{c|}{$\mathcal{B}(q_{min}^{2}\leq q^{2}\leq q_{max}^{2})$}\\
\cline{2-4}
&~~~This work~~~  &\cite{BPill-A.Ali-1312.2523} &\cite{BPill-W.S.Hou-1403.7410}\\
\hline
 $[0.05,2.0]$    &~~~~~~$0.177^{+0.021}_{-0.015}$~~~~~~   &~~~~~~$0.15^{+0.03}_{-0.02}$~~~~~~ &$--$\\
 $[1.0,2.0]$     &$0.092^{+0.010}_{-0.008}$   &$0.08^{+0.01}_{-0.01}$ &$--$\\
 $[2.0,4.3]$     &$0.215^{+0.021}_{-0.017}$   &$0.19^{+0.03}_{-0.02}$ &$--$\\
 $[4.3,8.68]$    &$0.426^{+0.037}_{-0.031}$   &$0.37^{+0.06}_{-0.04}$ &$--$\\
 $[0.05,8.0]$    &$0.751^{+0.070}_{-0.057}$   &$0.66^{+0.10}_{-0.07}$ &$--$\\
 $[1.0,6.0]$     &$0.470^{+0.045}_{-0.036}$   &$--$      &$0.44^{+0.06}_{-0.05}$\\
 $[2.0,6.0]$     &$0.378^{+0.035}_{-0.028}$   &$--$      &$0.36^{+0.05}_{-0.04}$\\
 $[1.0,8.0]$     &$0.666^{+0.060}_{-0.050}$   &$0.58^{+0.09}_{-0.06}$ &~~$0.63^{+0.09}_{-0.07}$~~~\\
 $[10.09,12.86]$ &$0.288^{+0.033}_{-0.025}$   &$0.25^{+0.04}_{-0.03}$ &$--$\\
 $[14.18,16.0]$  &$0.192^{+0.016}_{-0.014}$   &$0.15^{+0.03}_{-0.02}$ &$--$\\
 $[16.0,18.0]$   &$0.197^{+0.018}_{-0.016}$   &$0.15^{+0.03}_{-0.02}$ &$--$\\
 $[18.0,22.0]$   &$0.322^{+0.046}_{-0.042}$   &$0.25^{+0.04}_{-0.03}$ &$--$\\
 $[22.0,26.4]$   &$0.155^{+0.046}_{-0.039}$   &$0.13^{+0.02}_{-0.02}$ &$--$\\
 \hline
 $[12.0,14.0]$    &$0.218^{+0.018}_{-0.025}$   &$--$ &$--$\\
 $[14.0,16.0]$    &$0.212^{+0.017}_{-0.016}$   &$--$ &$--$\\
 $[16.0,18.0]$    &$0.197^{+0.018}_{-0.016}$   &$--$ &$--$\\
 $[18.0,20.0]$    &$0.176^{+0.021}_{-0.019}$   &$--$ &$--$\\
 $[20.0,22.0]$    &$0.147^{+0.025}_{-0.023}$   &$--$ &$--$\\
 $[22.0,24.0]$    &$0.106^{+0.027}_{-0.024}$   &$--$ &$--$\\
 $[24.0,26.4]$    &$0.049^{+0.019}_{-0.016}$   &$--$ &$--$\\
\hline
\end{tabular}
\end{table}

Neglecting the isospin-symmetry breaking effect, in the following
numerical discussion we take the $B^-\to\pi^-\ell^+\ell^-$ modes as
an illustrate example. In Fig.\ref{fig:dBSMeMu}, shown are the
dilepton invariant mass distributions in $B^-\to\pi^-\ell^+\ell^-$
$(\ell=e,\mu)$ obtained in the corresponding kinematically
accessible regions $4m^2_{\ell}\leq q^2\leq (m_B-m_{\pi})^2$ with
the predicted form factor shapes together with the parameter inputs
listed in Tab.~\ref{tab:InPut}, which coincide to a reasonably large
extent with each other. A similar distribution is given in
\cite{BPill-A.Ali-1312.2523}, however in the central value there
exists about a $-10\%$ to $-15\%$ deviation from the present
estimates, depending on $q^2$. Unfortunately, no experimental
measurement is available to make comparison. The branching ratios
are estimated at:
\begin{eqnarray}
\mathcal{B}(B^-\to\pi^-e^+e^-)&=&\left(2.263^{+0.172}_{-0.161}\big|_{\mathrm{CKM}}
~^{+0.143}_{-0.095}\big|_{\mathrm{FF}}~^{+0.040}_{-0.044}\right)\times10^{-8},\\\label{eq:Br-Mu}
\mathcal{B}(B^-\to\pi^-\mu^+\mu^-)&=&\left(2.259^{+0.172}_{-0.160}\big|_{\mathrm{CKM}}
~^{+0.141}_{-0.094}\big|_{\mathrm{FF}}~^{+0.040}_{-0.044}\right)\times10^{-8},
\end{eqnarray}
where the uncertainties originating from the related CKM matrix
elements and the form factors are separately given. The resulting
prediction \eqref{eq:Br-Mu} can be well accommodated by the
experimental result by the LHCb collaboration, but has a slightly
larger central value than those reported in
\cite{BPill-J.J.Wang-0711.0321,BPill-Z.J.Xiao-1207.0265,BPill-A.Ali-1312.2523,
Bpill-R.N.Faustov-1403.4466} (see Tab.\ref{tab:Br}). Also, it is of
specific experimental interest to compute the partial branching
ratios in some chosen $q^2$ intervals. In Tab.\ref{tab:PartialBr}
the results yielded for $B^-\to\pi^-\mu^{+}\mu^{-}$ are summarized
and some of them are compared with the predictions in
\cite{BPill-A.Ali-1312.2523} and in the context of QCDF
\cite{BPill-W.S.Hou-1403.7410}.
\begin{figure}[!ht]
\begin{center}
\includegraphics[scale=0.8]{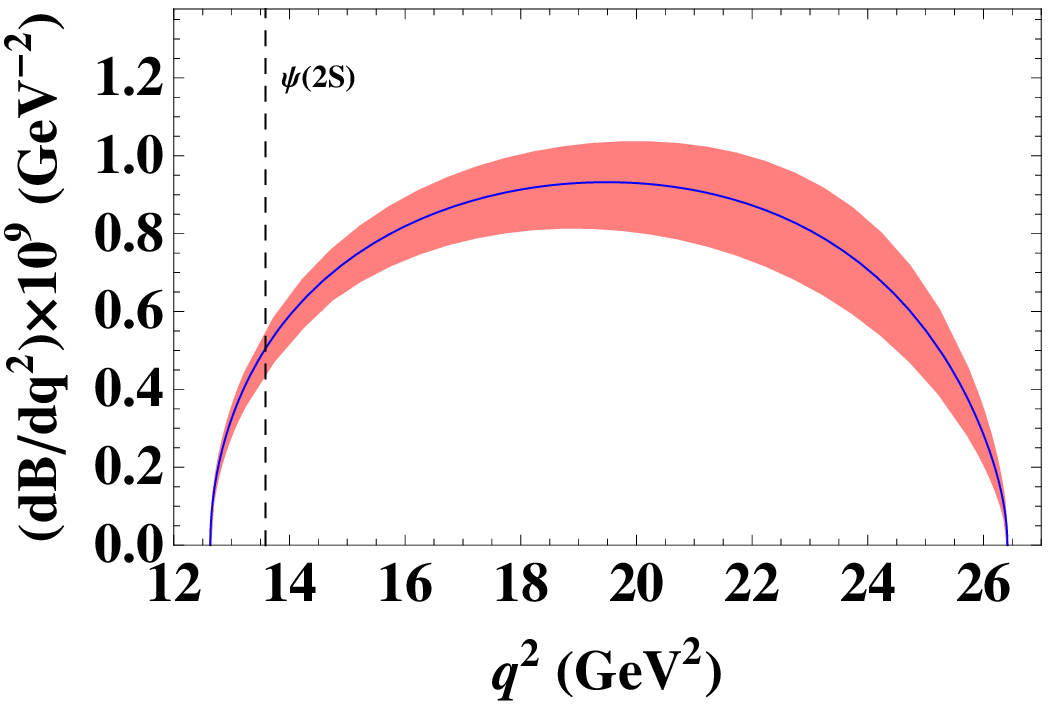}
\caption{Invariant mass distribution in
$B^-\to\pi^-\tau^{+}\tau^{-}$. The blue solid line indicates the
central values and the pink shadow region does the
uncertainties.}\label{fig:dBSMTau}
\end{center}
\end{figure}
\begin{table}[h]
\centering \caption{Theoretical predictions for the
$B^-\to\pi^-\tau^{+}\tau^{-}$ partial branching ratios in some
chosen $q^2$ intervals (in $10^{-8}$).}\label{tab:PartialBrTau}
\begin{tabular}{|c|c|}
\hline
 ~~~~$[q_{min}^{2},q_{max}^{2}]$~~~~ &~~~$\mathcal{B}(q_{min}^{2}\leq q^{2}\leq q_{max}^{2})$~~~\\
\hline
 $[12.6,14.0]$     &$0.056^{+0.005}_{-0.007}$    \\
 $[14.0,16.0]$     &$0.144^{+0.012}_{-0.015}$   \\
 $[16.0,18.0]$     &$0.175^{+0.016}_{-0.019}$   \\
 $[18.0,20.0]$     &$0.185^{+0.019}_{-0.023}$   \\
 $[20.0,22.0]$     &$0.182^{+0.023}_{-0.027}$   \\
 $[22.0,24.0]$     &$0.160^{+0.025}_{-0.029}$   \\
 $[24.0,26.4]$     &$0.114^{+0.022}_{-0.026}$   \\
\hline
\end{tabular}
\end{table}

It is absolutely essential to investigate the $B^-\to \pi^-
\tau^+\tau^-$ decay, which albeit is difficult to detect due to the
tau lepton's short lifetime. Other than the dielectronic and
dimuonic decays, this type of modes are made depend strongly on the
scalar form factor and have a much narrower kinematical region by
the large tau lepton mass. Hence they as well as $B\to \pi$
semileptonic decays into a tau lepton could be used to examine our
prediction for the form factor in intermediate and large $q^2$
region, under the prerequisite, of course, that all possible new
physics effects on these modes are negligibly small. We have the
invariant mass distribution shown in Fig.\ref{fig:dBSMTau} and the
branching ratio,
\begin{eqnarray}
\mathcal{B}(B^-\to \pi^-\tau^{+}\tau^{-})&=&\left(1.017^{+0.077}_{-0.072}\big|_{\mathrm{CKM}}
~^{+0.090}_{-0.119}\big|_{\mathrm{FF}}~^{+0.004}_{-0.006}\right)\times10^{-8}.
\end{eqnarray}
Compared with the previous studies using the pQCD method
\cite{BPill-Z.J.Xiao-1207.0265} and the relativistic quark model
\cite{Bpill-R.N.Faustov-1403.4466}, an increase of one order of
magnitude is here observed in the branching ratio, as shown in
Tab.\ref{tab:Br}. This difference is mainly caused by the difference
in the shape of the scalar form factor used as input. The partial
branching ratios assessed in several $q^2$ bins are collected in
Tab.\ref{tab:PartialBrTau}.

It is realistic to expect very soon the release of available LQCD
data on the $f_T^{B\to\pi}(q^2)$ form factor. Then we can have an
updated result for its $q^2$ behavior and thus for the observables,
which, however, is not expected to substantially improve the present
estimates.
\section{Summary}
We have reported a novel approach to the $B\to \pi\ell^+\ell^-$ form
factors in the whole semileptonic region, which combines the LCSR
method, LQCD simulation, $SU_F(3)$ symmetry breaking analysis and
form factor analyticity, and applied the resulting form factor
predictions to estimate several important observables of these rare
modes in naive factorization.

To twist-2 NLO accuracy and with the $\overline{\mathrm{MS}}$ mass
for the underlying $b$ quark, the shapes of the form factors in the
region $0\leq q^2 \leq 12~ \mathrm{GeV}^2$ are estimated in the LCSR
approach with a chiral current correlator, and a result free of
pollution by twist-3 is obtained. We further investigate their
behavior in the entire kinematically accessible region. For both
vector and scalar form factors, a simultaneous fit to a
two-parameter BCL series is carried out of the sum rule results in
an enlarged $q^2$ region and the corresponding LQCD ones available
at some high $q^2$'s. Given lack of available LQCD data on the
tensor form factor, as a similar procedure is applied for an
all-around understanding of its behavior we use as a constraint
condition at high $q^2$ the LQCD prediction for the corresponding
$B\to K$ form factor in conjunction with a $SU_F(3)$ symmetry
breaking ansatz. With the fitted parameterizations, we make a
prediction for the dilepton invariant mass spectra and branching
ratios for $B\to \pi\ell^+\ell^-$, by taking as an example the
$B^-\to \pi^-$ charged decay modes. For the dielectron and dimuon
modes, a branching ratio in good agreement with the experimental
measurement is obtained as,
$\mathcal{B}(B^-\to\pi^-e^+e^-)=(2.263^{+0.227}_{-0.192})\times10^{-8}$
and
$\mathcal{B}(B^-\to\pi^-\mu^+\mu^-)=(2.259^{+0.226}_{-0.191})\times10^{-8}$.
Consequently a more stringent constraint than what have been
achieved before is imposed on the possible new physics contribution.
In contrast, the corresponding observation made for the ditau mode
is
$\mathcal{B}(B^-\to\pi^-\tau^+\tau^-)=(1.017^{+0.118}_{-0.139})\times10^{-8}$,
which turns out to be one order of magnitude larger than the
previous predictions based on the form factor calculations by two
other approaches. We present also an assessment of the partial
branching ratios in some chosen $q^2$ bins, which can be confronted
with the future experimental data, along with the resulting
invariant mass distributions.

On the ground of the present findings and experimental measurement
for $B\to\pi\mu^+\mu^-$, it seems that there is less room left for
physics beyond the SM in the $B\to \pi$ dileptonic modes. However,
it is still too early to draw any final conclusion, because possible
new physics, if exists, would be expected to manifest itself through
some other observables, such as the CP and the forward-backward
asymmetries. Whereas in the SM a trustworthy result for the CP
asymmetry in the charged decay modes is difficult to achieve due to
our limited understanding of the two main sources of uncertainty,
the long-distance function $Y_{\mathrm{LD}}(q^2)$ and
weak-annihilation, the forward-backward asymmetry is precisely zero
and thereby its non-zero measurement would be a clean signal of new
effective couplings out of the SM scope.
\section*{ACKNOWLEDGEMENTS} N.~Zhu is indebted to
Dr.~Y.~-M.~Wang for enormously helpful discussion in the numerical
analysis of the sum rules. This work is in part supported by the
National Science Foundation of China (Key Program) under Grant Nos.
11235005, 11325525 and 11275114.



\end{document}